\documentclass[acmsmall]{acmart}

\usepackage[most]{tcolorbox}
\usepackage{listings}
\usepackage{graphicx}
\usepackage{colortbl} 
\usepackage{multirow}
\usepackage{amsmath}
\usepackage{mathtools}
\usepackage{subcaption}
\usepackage{balance}
\usepackage{color}
\usepackage{wrapfig}
\usepackage{float}
\usepackage[utf8]{inputenc}
\usepackage{cleveref}
\usepackage{xcolor}

\lstdefinestyle{custombox}{
    frame=single,                     
    framesep=1pt,                     
    rulecolor=\color{black},          
    numbers=left,                     
    numberstyle=\small\color{black},  
    numbersep=10pt,                   
    basicstyle=\fontsize{8.5}{10.5}\selectfont\ttfamily,       
    backgroundcolor=\color{gray!10},  
    breaklines=true,                  
    breakindent=0pt,                  
    xleftmargin=10pt,                 
    xrightmargin=10pt                 
}

\AtBeginDocument{%
  }

\setcopyright{cc}
\setcctype{by}
\acmJournal{PACMHCI}
\acmYear{2025} \acmVolume{9} \acmNumber{2} \acmArticle{CSCW161} \acmMonth{4} \acmDOI{10.1145/3711059}


\begin{document}

\title[Rideshare Transparency]{Rideshare Transparency: Translating Gig Worker Insights on AI Platform Design to Policy}

\author{Varun Nagaraj Rao}
\affiliation{%
  \institution{Center for Information Technology Policy, Princeton University}
  \country{USA}
  }
\email{varunrao@princeton.edu}

\author{Samantha Dalal}
\affiliation{%
  \institution{University of Colorado Boulder}
  \country{USA}
  }
\email{samantha.dalal@colorado.edu}

\author{Eesha Agarwal}
\affiliation{%
  \institution{Princeton University}
  \country{USA}
  }
\email{eagarwal@alumni.princeton.edu}

\author{Dana Calacci}
\affiliation{%
  \institution{Penn State University}
  \country{USA}
  }
\email{dcalacci@psu.edu}

\author{Andrés Monroy-Hernández}
\email{andresmh@princeton.edu}
\affiliation{%
  \institution{Center for Information Technology Policy, Princeton University}
  \country{USA}
  }

\renewcommand{\shortauthors}{Varun Nagaraj Rao, Samantha Dalal, Eesha Agarwal, Dana Calacci, \& Andrés Monroy-Hernández}

\begin{abstract}
Rideshare platforms exert significant control over workers through algorithmic systems that can result in financial, emotional, and physical harm. What steps can platforms, designers, and practitioners take to mitigate these negative impacts and meet worker needs?
In this paper, we identify transparency-related harms, mitigation strategies, and worker needs while validating and contextualizing our findings within the broader worker community.
We use a novel mixed-methods study combining an LLM-based analysis of over 1 million comments posted to online platform worker communities with semi-structured interviews with workers.
Our findings expose a transparency gap between existing platform designs and the information drivers need, particularly concerning promotions, fares, routes, and task allocation.
Our analysis suggests that rideshare workers need key pieces of information, which we refer to as \textit{indicators}, to make informed work decisions.
These indicators include details about rides, driver statistics, algorithmic implementation details, and platform policy information. 
We argue that instead of relying on platforms to include such information in their designs, new regulations requiring platforms to publish public transparency reports may be a more effective solution to improve worker well-being. 
We offer recommendations for implementing such a policy.
 \end{abstract}


\received{January 2024}
\received[revised]{July 2024}
\received[accepted]{October 2024}

\maketitle

\section{Introduction}

Rideshare platforms match people seeking transportation (riders) with transportation providers (drivers). They do so by constructing algorithmically managed ``marketplaces''~\cite{uber2023marketplace} where they match riders' demand with drivers' supply. While algorithmic management enables rideshare platforms to operate at scale, it can cause workers financial, emotional, and physical harm through gamification, omnipresent surveillance, and opaque algorithmic decisions \cite{woodcockGamificationWhatIt2018, dubal2023algorithmic, dziezaRevoltDeliveryWorkers2021, figueroaEssentialUnprotectedAppbased, rosenblat2016algorithmic, shapiroAutonomyControlStrategies2018a, watkinsFaceWorkHumanCentered2023}. Workers, in an attempt to regain control and develop infrastructural competency and coping mechanisms, engage in resistance strategies (e.g., canceling rides), switching apps (e.g., transitioning from Uber to Lyft), and gaming the system (e.g., activating the app once inside surge regions) \cite{mohlmann2017hands, lee2015working}.

This dynamic also subjects drivers to \textit{information asymmetry}, i.e., an imbalance in the information accessible to rideshare platforms and drivers~\cite{rosenblat2016algorithmic}. As digital environment designers, rideshare platforms create an asymmetric information environment and then exploit it using mechanisms of ``soft control'' \cite{viljoenDesignChoicesMechanism2021, tomassetti2016does, deleuze1995postscript} that frequently disadvantage workers and prevent them from making fully informed decisions \cite{shapiroAutonomyControlStrategies2018a, rosenblat2016algorithmic}. 
For instance, platforms manipulate driver supply using surge pricing bonuses that increase driver pay during periods of high customer demand \cite{shapiroAutonomyControlStrategies2018a}.
Concretely, this asymmetry manifests to drivers through design choices that impact the \textit{transparency}—the understandability and accessibility—of important information \cite{bommasani2023foundation}. 

Although transparency in platform design is key to how drivers experience information asymmetry and its harms, its role in shaping working conditions and workers' coping strategies is under-studied. Previous research on algorithmic management has examined the harms and associated mitigation strategies used by drivers around finances~\cite{watkins2022have}, physical safety~\cite{qadriSeeingDriverHow2022}, and social connection~\cite{yao2021together}, but has not explored these same issues around transparency or how it manifests in platform interface design. Other work that examines worker \textit{needs} due to algorithmic management uses speculative design scenarios and manufactured exercises rather than evaluating existing platform design \cite{hsieh2023co, alvarez2022design, zhang2022algorithmic, zhang2023stakeholder}.

In this paper, we investigate how a lack of transparency in platform design, particularly regarding AI and algorithmic decisions, impacts drivers' working conditions and coping strategies. Specifically, our research questions are \textit{what are the transparency-related harms, mitigation strategies, and needs of rideshare workers?}, and \textit{what policy solutions may mitigate worker harms and address workers' needs?} Our contributions are as follows:
\begin{enumerate}
    \item We comprehensively identify rideshare workers' transparency concerns, mitigation strategies, and needs
    \item We obtain empirical evidence from workers through a unique mixed methods study blending in-depth interviews and an LLM-based analysis of rideshare community discussions on Reddit
    \item We offer a detailed roadmap for a \textit{rideshare transparency report}, a policy solution that we argue could address driver transparency needs and mitigate some of the harms we identify.
\end{enumerate}

To uncover transparency-related harms, mitigation strategies, and needs of rideshare workers, we conducted in-person interviews with (N=9) Uber and Lyft drivers in the Northeast United States. Drawing on recent methods in human-centered explainable AI \cite{ehsan2020human, kim2023help, suresh2021beyond}, we developed an annotation-based activity. This activity involved workers annotating screenshots of app UIs during interviews, providing insights into worker-generated design improvements based on their real interactions with the platform apps they work for. To validate findings on a larger scale, we analyzed over 1 million submissions and comments from Uber and Lyft driver discussions on Reddit. Using a novel LLM-based method, taking a step further than traditional topic modeling \cite{gillies2022theme, abdul2018trends, cambo_ModelPositionalityComputational_2022, muller_MachineLearningGrounded_2016a, baumer_ComparingGroundedTheory_2017a}, we extracted transparency-related harms. The capabilities of LLMs, beyond conventional topic modeling, allowed for more nuanced and contextually relevant identification of these harms in the dataset. Our analysis confirmed that transparency harms identified from Reddit data corroborate with interview findings.

Our results highlight a transparency gap between the information offered by existing platform designs and the information drivers believe they need to make informed decisions and maintain their well-being. 
Specifically, we find that drivers experience transparency-related harms when utilizing four key features of rideshare platforms: promotions, fares, routes, and task allocation, prompting a desire for more information in these contexts.
Beyond the design recommendations made by drivers we document in our findings, we argue for a policy solution to close this gap. 
We propose regularly-released rideshare transparency reports, drawing inspiration from successful precedents by social media platforms and AI model regulation \cite{Heikkilä2023high, diresta2022time, bommasani2023foundation, edelson2021standard, silverman2023tiktok}. These reports would encompass crucial indicators such as individual ride statistics, driver history, and details about platform algorithms' inputs and outputs. Access to this information  could improve worker well-being and address major harms.
We offer a detailed roadmap for what content such reports should include and some strategies for how regulatory agencies could implement them, contributing to the ongoing discourse on rideshare platform regulation and design.
We note here that this research is situated within the U.S. context, drawing solely upon data from English-language forums on Reddit and driver interviews in the North-East U.S. While the findings provide structural insights into the rideshare industry that can impact any driver, the policy contributions are primarily geared towards the U.S. setting.

\section{Related Work}
\label{sec:related-work}

\subsection{The Rideshare Economy and its Impacts on Workers}

Rideshare platforms and their impacts on workers have been well documented within the CSCW literature. While some CSCW research on rideshare platforms propose design recommendations to improve outcomes for workers, few papers take into account the feasibility of their recommendations given the existing legal regimes platforms operate within and how they shape platforms' operating logics~\cite{dalal2023understanding}. In this paper, we show how empirical studies of rideshare platform features can inform targeted policy solutions that can potentially alter the legal regimes platforms operate within. To situate our solution, we first provide a brief summary of the existing state of CSCW and HCI research on rideshare platforms, how these fields approach improving outcomes for rideshare workers, and precedents for using policy to increase transparency in other fields.

\subsubsection{The Rideshare Economy}
The rideshare economy, a prominent sector of the larger gig economy, has become ingrained in the transportation and mobility landscape of the U.S. In the U.S. alone, over 36\% of adults have used rideshare platforms, and an estimated 5\% of the American workforce has driven for a rideshare platform for work~\cite{pewresearch}. Despite the rideshare economy's established presence in the transportation sector, there is a lack of detailed rideshare driver workforce data because it is difficult to track these workers due to their classification as independent contractors~\cite{abrahamMeasuringGigEconomy2018}. While the Bureau of Labor Statistics made some initial strides in documenting key information about the gig economy~\cite{bls2017contingent}, reliable documentation of working conditions for rideshare workers, such as rate of workplace injuries, average wages, and demographic make-up remains difficult as there is no stable mechanism to systematically track independent contractors in the gig economy~\cite{zipperer2022national,donovan2016does,abrahamMeasuringGigEconomy2018}. A lack of data about labor conditions in the rideshare industry contributes to regulatory shortcomings \cite{datadeficit}. However, researchers have chipped away at this lack of visibility into rideshare working conditions through rich qualitative studies with drivers to understand their experiences and the harms they face. 

\subsubsection{Known Harms and Mitigation Strategies in Rideshare Work}
Prior research extensively documents that rideshare workers face financial, psychological, and physical harm due to opaque algorithmic decisions, surveillance, and hazardous conditions. Rideshare platforms' gamification leads to long work hours and uncertain profits, pushing drivers to work faster and harder, which can result in physical harm \cite{dziezaRevoltDeliveryWorkers2021,figueroaEssentialUnprotectedAppbased, krzywdzinskiAutomationGamificationForms2021}. Omnipresent surveillance causes psychological harm \cite{rahmanInvisibleCageWorkers2021a, shapiroAutonomyControlStrategies2018a}. Deactivation \footnote{In gig work, deactivation is the same as being fired. Workers lose access to their accounts and cannot earn money on the platform until the account is reinstated.} without recourse results in financial harm \cite{watkinsFaceWorkHumanCentered2023, schwartzDeactivationRepresentationRole2023}. Our large-scale analysis of over one million Reddit comments, combined with interview data, reveals the persistence of these harms at scale, surpassing the scope of previous small-scale studies ~\cite{rosenblat2016algorithmic, sannon2022privacy, yao2021together, watkins2022have, ma2018using}. While the harms of the rideshare industry have been well-documented in qualitative work, few studies have explored what \textit{specifically} should be changed within platform interfaces to mitigate harms and how policy can be used as a tool to enforce changes to platform interfaces. Our study seeks to mitigate \textit{information asymmetries} in rideshare work by identifying features within the platforms needing change and calling for policymakers to enforce public data disclosures.

\subsection{Calls for Transparency}
\subsubsection{Transparency Requirements in Rideshare Industry:}
Researchers, journalists, and worker advocates highlight that the opaque nature of algorithmic management in gig work causes financial, physical, and psychological harm due to a lack of transparency, increasing the precarity of such work \cite{dubal2023algorithmic, uber2019forbes, uber2023guardian}. Despite this, prior research lacks proposals for concrete solutions to enhance transparency. Responding to this gap, rideshare driver organizations, backed by legislative efforts like Colorado's SB24-75 \cite{colorado2024transparency} and Washington State's House Bill 2076 \cite{wa2022transparency}, demand clearer insights into task allocation, earnings, and deactivation processes. The need for transparency is echoed in robust discussions on driver forums like Reddit and UberPeople. While prior research demonstrates that lack of transparency exacerbates the harms rideshare drivers face~\cite{rahmanInvisibleCageWorkers2021a,shaikhWorkMakePiecework,woodcock2020algorithmic,qadriSeeingDriverHow2022,zhang2023stakeholder,yao2021together,watkins2022have}, the specific transparency harms that workers face and their needs for increased information access are under-explored.

Policy solutions are essential to address the needs of rideshare workers effectively, especially since platforms typically lack the incentive to prioritize these needs over their profit-maximization objectives~\cite{zuboff2015big}. Prior policy recommendations made in CSCW research have been broad in their specifications and not implemented into policy practice \cite{hsieh2023designing, van2023migration}. We propose a novel and actionable policy solution: advocating for a rideshare transparency report that provides specific guidance on what data should be made transparent. This policy lever draws inspiration from their impact in other digital technologies such as AI models and social media \cite{Heikkilä2023high, diresta2022time, bommasani2023foundation, edelson2021standard, silverman2023tiktok}. We believe such specific reports could lead to concrete legislative actions, similar to the EU's Digital Services Act and AI Act, and the US's proposed Platform Accountability and Transparency Act, thereby concretely addressing some needs of rideshare workers.

\subsubsection{Transparency Requirements in Other Industries:} Transparency data disclosures are mandated in several industries. For food and drugs, the Nutrition Labeling and Education Act of 1990 mandates nutritional labeling, while the FDA's FAERS\footnote{FDA Adverse Event Reporting System: \url{https://catalog.data.gov/dataset/fda-adverse-event-reporting-system-faers-latest-quartely-data-files}} requires reporting of adverse drug events. Meanwhile, for AI models, datasheets\cite{gebru2021datasheets}, model cards\cite{mitchell2019model}, and transparency indices\cite{bommasani2023foundation} have been proposed, though no legal mandates exist yet aside from draft provisions in regulations like the EU AI Act and US AI Foundation Model Transparency Act. In the corporate realm, 154 Fortune 500 companies voluntarily release diversity reports\footnote{Diversity data in Purpose Brand’s 2023 survey of corporate DEI reports found 154 published diversity reports from Fortune 500 companies: \url{https://purposebrand.com/blog/diversity-report-examples-fortune-2023/}}, and Title VII requires affirmative action plans\footnote{Establishing affirmative action plans: \url{https://www.law.cornell.edu/cfr/text/29/1608.4}}. Financial reporting is stringently regulated, with the Sarbanes-Oxley Act of 2002 mandating that public companies file annual (10-K), quarterly (10-Q), and event-based (8-K) reports with the U.S. Securities and Exchange Commission to provide insights into a company's financial health. Additionally, 88 telecommunications and social media companies\footnote{AccessNow's 2021 index features a record of transparency reports from leading internet companies and telcos: \url{https://www.accessnow.org/campaign/transparency-reporting-index/}} provide transparency reports on user data and content moderation, driven by privacy and surveillance concerns from the FBI\cite{fbi2010aclu} and NSA\cite{NSA2013verizon}, while the EU's Digital Services Act mandates such reporting for very large online platforms\footnote{The EU's DSA Transparency Database: \url{https://transparency.dsa.ec.europa.eu/}}. However, binding U.S. laws are still lacking in this domain and there could be tensions with the First Amendment due to concerns of compelled speech.

\subsection{Design Research for Rideshare Work}
Worker-centered design research has uncovered workers' imaginaries for new systems of work (~\cite{zhang2022algorithmic,bates2021lessons}, led to the creation of tools to support worker self-reflection and well-being (~\cite{you2021go,zhang2023stakeholder}), and enabled collective action through information sharing (~\cite{irani2013turkopticon,salehi2015we,calacci2022bargaining}). However, as ~\citet{zhang2023stakeholder} points out, incorporating end-users into the re-design of algorithmic systems can be fraught due to AI systems' dynamic nature: these systems are often unstable and opaque which complicates designers' approaches to prototyping. As a result, we observe that most prior gig work centered design research has utilized speculative scenarios to elicit workers needs. For example, ~\citet{zhang2023stakeholder} utilize data-probes to familiarize rideshare workers with how algorithmic management systems function and enable them to make design recommendations. And, ~\citet{hsieh2023co} identified harms that gig workers face and utilized a speed-dating approach to efficiently surface workers' ideas for mitigating harms through speculative design. 

Human-centered explainable AI (HCXAI) research~\cite{ehsan2020human} however, highlights the importance of basing AI harm analysis and design recommendations on real user interactions, like those experienced by rideshare drivers (a group of AI end-users), instead of hypothetical scenarios ~\cite{tonekaboni2019clinicians, cai2019hello, cai2019human}. This aligns with recent trends in XAI that cater to specific user needs \cite{kim2023help, suresh2021beyond}. Our study focuses on the tangible transparency and explainability needs of rideshare workers, emphasizing the importance of real-world instances of harm for effective, user-centric solutions. We utilize 10 screenshots (detailed in Appendix \ref{asec:protocol-screenshots}) comprehensively encompassing the AI and algorithmic decisions on rideshare platforms to ground our in-person interviews with real-world worker needs and surface contextualized design features. This approach is unique in gig worker research for using such comprehensive, real-world instances to identify and ground algorithmic harm.

\section{Background on Rideshare Platform Features}
\label{sec:background}

\begin{figure}[htb]
    \centering    \includegraphics[width=\columnwidth]{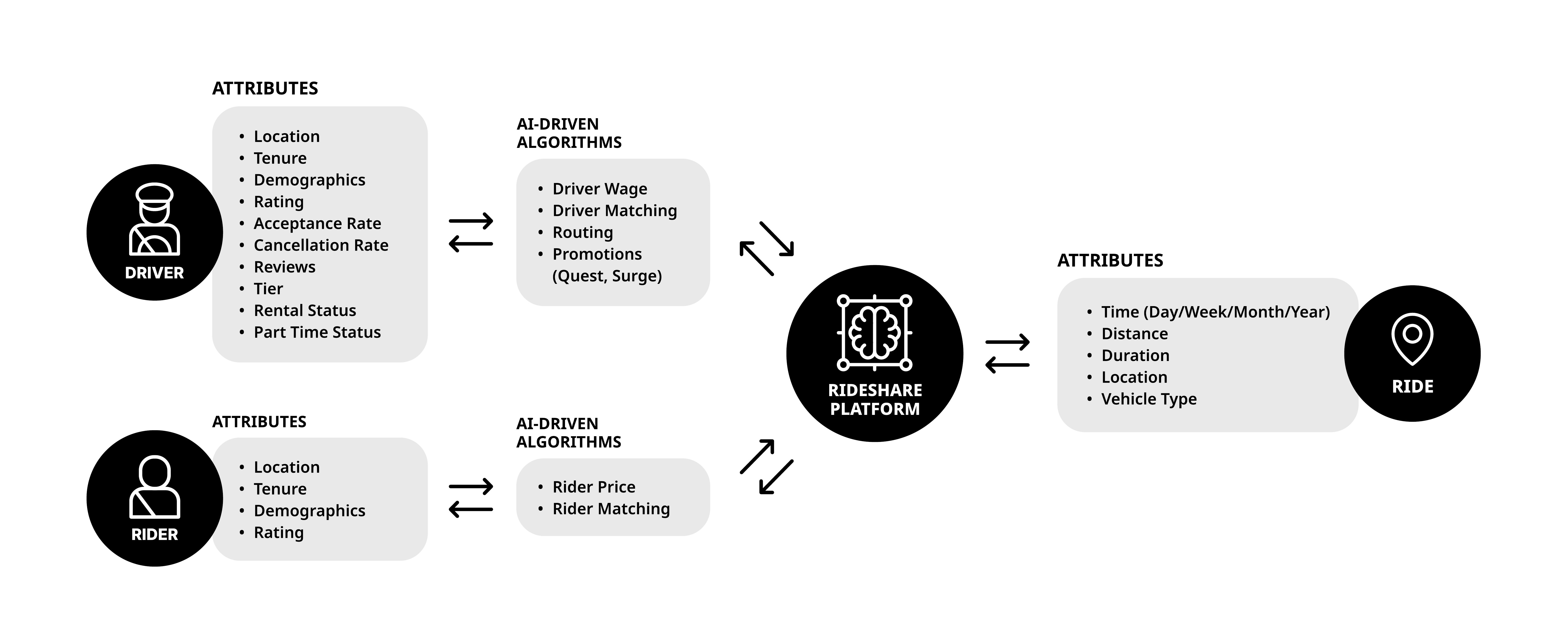}
    \caption{Algorithmic Landscape of Rideshare Platforms described using Uber's terminologies}
    \label{fig:rideshare-ecosystem}
\end{figure}

We provide a brief overview of the key AI and algorithmic aspects of rideshare platforms to help the reader understand the vocabulary participants used in our interviews, with a focus on Uber, the largest rideshare platform in the world (see Figure \ref{fig:rideshare-ecosystem} for a schematic representation). These insights also apply broadly to similar platforms such as Lyft and Ola, among others, but may be termed differently.

\subsection{AI and Algorithmic Features}

\subsubsection{Driver-Rider Matching}
Uber's algorithm for matching drivers and riders goes beyond simple proximity. It also integrates traffic data, geographical considerations, and advanced AI techniques, aiming to streamline wait times, optimize driver earnings, and enhance safety by preventing matches with a history of negative interactions \cite{uber2023match, uber2017ml, uber2018ml}.

\subsubsection{Driver Wages and Rider Prices}
Uber employs two distinct pricing models. The ``upfront pricing'' model provides both drivers and riders with predetermined earnings and costs for a trip, including detailed route information and estimated time \cite{uber2023upfront} based on several attributes like the time, distance, duration, location and vehicle types of the ride. Alternatively, in ``rate card'' regions, driver earnings are based on distance and time, revealed after the ride's completion. Across both pricing models, the platform takes a variable percentage of the rider's payment (colloquially referred to as the take rate). This pricing structure has been scrutinized for potential discriminatory practices \cite{pandey2021disparate, chen2015peeking, yan2020dynamic}.

\subsubsection{Routing}
The platform's AI-driven routing system calculates the most efficient paths to destinations. It also allows drivers the flexibility to use external navigation tools like Google Maps, Waze, or Garmin for guidance \cite{uber2018ml, uber2017ml}.

\subsubsection{Gamified Promotions and Algorithmic Features}
Uber implements various gamified incentives to regulate demand and motivate drivers. Surge pricing temporarily increases earnings in high-demand areas to attract more drivers \cite{uber2023surge}. Boosts offer additional pay for trips originating from specific zones during designated times \cite{uber2023boost}. Quests provide additional earnings for achieving certain rides or earnings milestones, e.g., drive 10 trips and get an extra \$25 within a set period \cite{uber2023quests}. 
Trip Radar displays nearby trip requests, allowing drivers to select multiple they're interested in, with matching prioritizing proximity and reduced wait times \cite{uber2022radar}. 

\subsection{Stakeholder Attributes}

\subsubsection{Driver Attributes}
Drivers on the platform can be characterized by their location, tenure, demographics, ratings, reviews, acceptance and cancellation rates, and their tier status (e.g., Uber Blue, Gold, Platinum, Diamond). The influence of these attributes on the platform's algorithms is a subject of ongoing driver concern and debate. These tiers, linked to performance metrics such as acceptance and cancellation rate, offer various rewards and benefits \cite{uber2023pro}. Drivers' ability to rate and review riders is also a key aspect of the platform's feedback system \cite{uber2023rating}.

\subsubsection{Rider Attributes}
Riders can be characterized by their location, tenure, demographics, ratings, and reviews. Similar to driver attributes, the influence of these rider attributes on the platform's algorithms is a subject of ongoing debate. Riders, like drivers, have the capability to rate and review drivers \cite{uber2023rating}.

\section{Methods}
\label{sec:methods}

\begin{figure}[htb]
    \centering    \includegraphics[width=\columnwidth]{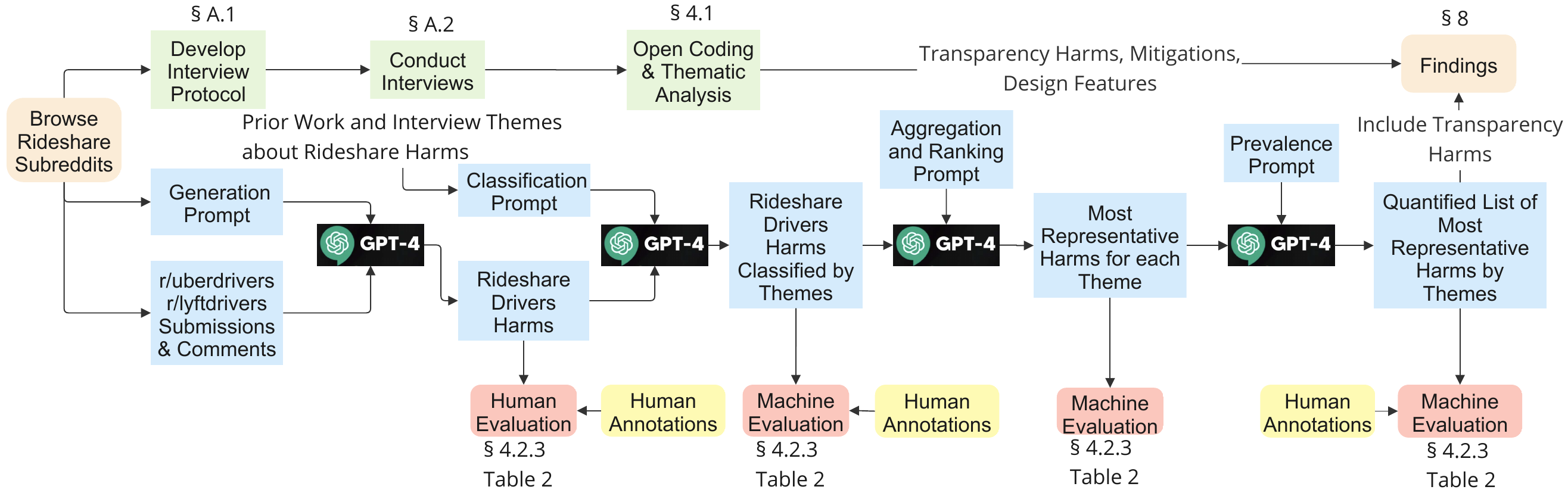}
    \caption{Overview of our Methods}
    \label{fig:methods-flowchart}
\end{figure}

We used a mixed-methods approach to answer our research questions.
We first browsed Reddit for a broad understanding of discussions of rideshare workers relevant to the platforms' non-transparent AI and algorithmic decisions. We then developed an interview protocol and conducted interviews to surface drivers' concerns, mitigation strategies, and needs relevant to transparency. In parallel, to complement the depth of interviews with the breadth of other drivers' concerns \footnote{This analysis was primarily limited to surfacing harms, as most Reddit discussions centered on voicing concerns \cite{yao2021together, watkins2022have}.} across a larger population, we used a Large Language Model (LLM) to analyze all \texttt{r/uberdrivers} and \texttt{r/lyftdrivers} posts (65K) and comments (1.4MM) from 2019-2022 on Reddit, leveraging initial broad themes from our interviews and prior work described in Section \ref{sec:related-work}. Finally, we compiled repeating themes from our interviews into a codebook, and incorporated relevant data from the Reddit analysis, resulting in four key transparency needs: promotions, fares, routes, and task allocation, to frame our findings sections. See Figure \ref{fig:methods-flowchart} for a visual overview of our methods.

\subsection{Method 1: Interviews with drivers}

We conducted semi-structured interviews with nine Uber and Lyft drivers (see Table \ref{tab:demo-info}
\begin{table}[tb]
\small
\resizebox{\textwidth}{!}{%
\begin{tabular}{@{}ccccllclc@{}}
\toprule
\textbf{P} &
  \textbf{Age} &
  \textbf{Race} &
  \textbf{Gender} &
  \textbf{Platforms} &
  \textbf{\# of Rides} &
  \textbf{Tenure} &
  \multicolumn{1}{c}{\textbf{Education}} &
  \textbf{Full Time} \\ \midrule
1 & 26 & White    & Male   & Lyft, Uber & 12,000 (Lyft) & 5 years  & Some college, no degree          & Yes \\
2 & 35 & Asian    & Male   & Uber       & 500 (Uber) & 2 months & Bachelor's degree                & Yes \\
3 & 55 & Black    & Female & Lyft, Uber & 600 (Uber) & 6 years  & High school degree or equivalent & No  \\
4 & 31 & Black    & Male   & Uber, Lyft & 4,300 (Uber) & 4 years  & Master's degree                  & No  \\
5 & 31 & Black    & Female & Uber       & 3,200 (Uber) & 6 years  & Some college, no degree          & No  \\
6 & -  & Black    & Male   & Lyft, Uber & 2,900 (Lyft) & 6 years  & Some college, no degree          & No  \\
7 & -  & Black    & Male   & Uber       & 23,000 (Uber) & 7 years  & Associate degree                 & No  \\
8 & 49 & White    & Male   & Uber, Lyft & 3,900 (Uber) & 8 years  & Associate degree                 & No  \\
9 & 36 & Hispanic & Male   & Uber, Lyft & 11,200 (Uber) & 5 years  & Some college, no degree          & Yes \\ \bottomrule
\end{tabular}%
}
\caption{Participant demographic and background information}
\label{tab:demo-info}
\end{table}
 for participant's demographics). We used \textit{non-probabilistic purposive sampling} \cite{patton2002two,crouch2006logic,o2013unsatisfactory} to recruit participants. The goal of the interviews is \textbf{not} to create generalizable findings, but rather to investigate information-rich cases that provide a ``thick description'' \cite{geertz2008thick} of transparency in rideshare work. The interviews involved participants annotating and discussing screenshots of the ridesharing apps printed on paper. The screenshots centered around the platform's AI and algorithmic decisions across various stages of a trip. We gathered the screenshots from discussions in the \texttt{r/uberdrivers} and \texttt{r/lyftdrivers} subreddits, which we found searching for phrases such as ``surge pricing,'' ``take rate,'' and ``upfront price,'' among others. For each screenshot, participants identified their specific transparency concerns with the information presented, described their current strategies for mitigating their concerns, and suggested design interventions to address them. The interview protocol and screenshots are available in the Appendix \ref{asec:protocol-screenshots}. The study was approved by our university IRB. 

\subsubsection{Participant Recruitment:} We recruited drivers who were active on Uber or Lyft in the last two years. We confirmed their activity through a screening survey that included uploading a driver profile screenshot. We recruited part-time and full-time drivers but excluded those solely providing delivery services. We posted recruitment messages on several NJ/NY Uber and Lyft driver groups on Facebook and WhatsApp, as well as on university Slack groups. Also, we paid to post recruitment messages on Craigslist and Reddit (as ads on \texttt{r/uberdrivers} and \texttt{r/lyftdrivers}). %

\subsubsection{Interview Process:} The interviews took place in person between November to December 2023. The interviews began with participants signing a consent form and completing a demographic survey, followed by an ice-breaker activity. During the interview, we examined drivers' transparency concerns throughout the three stages of a ride - at or before \textit{pickup} (Screenshots 1-5), during the \textit{ride} (Screenshot 6), at or after \textit{dropoff} (Screenshots 7-10), using 10 Reddit-sourced screenshots (see Appendix \ref{asec:protocol-screenshots} for the screenshots used). The screenshots included scenarios about ride matching, wage calculations, routing, promotions (quests, boosts, surge, trip radar), acceptance and cancellation rates, ratings and reviews, and tiers. Participants identified their concerns in these screenshots, discussed current mitigation strategies, and suggested additional desired information. 

The lead author conducted the interviews in person, which lasted between 90-100 minutes. We recorded interviews for transcription purposes. We compensated participants with an \$80 bank transfer or Amazon gift card. The first and second authors coded the transcriptions using thematic analysis. We conducted preliminary coding where the two researchers independently analyzed the same batch of interviews. We then convened to compare and discuss our findings. We also conducted card-sorting exercises to arrive at the final codes. Later, we compiled repeating themes in a codebook, which we then used to code the remainder of the interviews. We identified four key transparency needs that we use to frame our findings: promotions, fares, routes, and task allocation. 

\subsection{Method 2: Analysis of Online Discussion Data}

We analyzed large-scale text data from a large population of rideshare drivers in online communities to complement our interviews. 

\subsubsection{Dataset Summary:} 
We downloaded posts and comments from \texttt{r/uberdrivers} and \texttt{r/lyftdrivers} for 2019-2022 from the Reddit Pushift Archives \footnote{Reddit Pushift Archives: \url{https://the-eye.eu/redarcs/}}. 
After preprocessing to exclude the bottom 10\% of submissions and comments by word count for being too brief, our dataset included 65,377 submissions ($47,106$ from Uber, $18,271$ from Lyft) and $1,392,776$ associated comments ($1,054,030$ from Uber, $338,746$ from Lyft). Refer to Appendix \ref{asec:summary-stats} for detailed dataset summary statistics.

\subsubsection{Multi-phase LLM Prompting for Analysis:} 
We used an LLM (GPT4-Turbo OpenAI model) to extract a quantified list, based on the frequency of occurrence of rideshare workers' concerns, from Reddit \footnote{Refer to the Appendix \ref{asec:llm-background} for more context on the use of LLMs for analysis of public forum data}. We used the following four prompts to guide our analysis (see Appendix \ref{asec:prompt} and \ref{asec:costs} for prompts details and associated costs). 
\begin{enumerate}
    \item \textbf{Generation Prompt:} Identified posts and comments in our dataset that described concerns surrounding platforms' AI and algorithmic decisions. 
    \item \textbf{Classification Prompt:} Using four commonly highlighted rideshare harms' themes (transparency, predictability, safety and fairness) from prior work \cite{rosenblat2016algorithmic, zhang2022algorithmic, hsieh2023designing, wa2022transparency, hess2019transparency} and our own interviews, we classified each output concern from the generation prompt into one of four primary themes, with an additional ``Other'' category. 
    \item \textbf{Aggregation Prompt:} We aggregated and obtained the five most frequently occurring concerns (secondary themes) within each primary theme.
    \item \textbf{Prevalence Prompt:} We classified concerns within each primary theme as belonging to one of the secondary themes or an ``Other'' category. 
\end{enumerate}

Subsequently, the researchers selected all transparency-related concerns, grouped them according to the four interview codes: promotions, fares, routes, and task allocation, and quantified their prevalence through raw counts and a percentage value across all the transparency-related concerns. We present the quantitative analysis in Tables \ref{tab:reddit-promotions}, \ref{tab:reddit-fare}, \ref{tab:reddit-routing}, \ref{tab:reddit-task}, of the findings section.

\subsubsection{LLM Performance Evaluation Metrics}
We define metrics across all analysis phases to evaluate the LLM outputs' performance. Table \ref{tab:llm-evaluation} presents the metrics, calculation methods, reference data generation methods, and results. All metrics range from 0 to 1, with higher values indicating better performance. Performance is generally very high, mostly exceeding 0.74. The low factuality score (0.55) is less concerning because the first analysis phase (Generation) uses broad instructions to allow ambiguity and flexibility. As the completeness score is quite high, most possible concerns move to the next stages, where more precise prompts, based on our interview analysis and motivated by prior work, exclude non-factual outputs. Additional details are available in Appendix \ref{asec:evaluation}. 
\begin{table}[]
\resizebox{\textwidth}{!}{%
\begin{tabular}{@{}l|lllll@{}}
\toprule
\rowcolor{black}
\color{white}\textbf{Evaluation Metrics} &
  \multicolumn{1}{c}{\cellcolor{black}\color{white}\textbf{Factuality}} &
  \multicolumn{1}{c}{\cellcolor{black}\color{white}\textbf{Completeness}} &
  \multicolumn{1}{c}{\cellcolor{black}\color{white}\textbf{Distinctness}} &
  \multicolumn{1}{c}{\cellcolor{black}\color{white}\textbf{Coverage (k)}} &
  \multicolumn{1}{c}{\cellcolor{black}\color{white}\textbf{Accuracy}} \\ \midrule
\textbf{Definition} &
  \begin{tabular}[c]{@{}l@{}}Proportion of LLM-\\ generated concerns \\ also identified by \\ humans\end{tabular} &
  \begin{tabular}[c]{@{}l@{}}Proportion of human-\\ identified concerns \\ also generated by the\\ LLM\end{tabular} &
  \begin{tabular}[c]{@{}l@{}}Proportion of LLM-\\ generated themes \\ where the most similar \\ topic to each theme is \\ unique compared to the \\ most similar topics of \\ the other themes.\end{tabular} &
  \begin{tabular}[c]{@{}l@{}}Proportion of the `n' LLM-\\ generated themes whose \\ most similar topic is \\ among the top `nk' most \\ frequent topics in the \\ overall text.\end{tabular} &
  \begin{tabular}[c]{@{}l@{}}Proportion of LLM \\ labels which match \\ human labels\end{tabular} \\ \midrule
\textbf{\begin{tabular}[c]{@{}l@{}}Analysis phase \\ evaluated with \\metric\end{tabular}} &
  Generation &
  Generation &
  Aggregation &
  Aggregation &
  \begin{tabular}[c]{@{}l@{}}Classification, \\ Prevalence\end{tabular} \\ \midrule
\textbf{\begin{tabular}[c]{@{}l@{}}Metric \\ calculation \\ method\end{tabular}} &
  \begin{tabular}[c]{@{}l@{}}Manually compare \\ reference human \\ concerns with \\ LLM-generated \\ concerns for a \\ sample of input \\  data\end{tabular} &
  \begin{tabular}[c]{@{}l@{}}Manually compare \\ reference human \\ concerns with \\ LLM-generated \\ concerns for a \\ sample of input \\  data\end{tabular} &
  \begin{tabular}[c]{@{}l@{}}Automatically determine \\ whether most similar \\ topic to each theme is \\ unique using topic \\ models\end{tabular} &
  \begin{tabular}[c]{@{}l@{}}Automatically compare \\ the most similar topic to \\ each theme with the most \\ frequently occurring \\ topics across the entire \\ text using topic models.\end{tabular} &
  \begin{tabular}[c]{@{}l@{}}Automatically \\ compare the LLM \\ labels with \\ human labels for a \\ sample of input \\  data\end{tabular} \\ \midrule
\textbf{\begin{tabular}[c]{@{}l@{}}Reference data\\ generation \\ method\end{tabular}} &
  \begin{tabular}[c]{@{}l@{}}Theme extraction \\ by human \\ (2 researchers)\end{tabular} &
  \begin{tabular}[c]{@{}l@{}}Theme extraction \\ by human \\ (2 researchers)\end{tabular} &
  \begin{tabular}[c]{@{}l@{}}Topic extraction using \\ topic models \\ (BERTopic)\end{tabular} &
  \begin{tabular}[c]{@{}l@{}}Topic extraction using \\ topic models \\ (BERTopic)\end{tabular} &
  \begin{tabular}[c]{@{}l@{}}Labeling \\ by humans \\ (2 researchers)\end{tabular} \\ \midrule
\textbf{\begin{tabular}[c]{@{}l@{}}Comparison b/w \\ reference data and \\ LLM output\end{tabular}} &
  \begin{tabular}[c]{@{}l@{}}Manually \\ (2 researchers)\end{tabular} &
  \begin{tabular}[c]{@{}l@{}}Manually \\ (2 researchers)\end{tabular} &
  \begin{tabular}[c]{@{}l@{}}Automatically \\ (Python Program)\end{tabular} &
  \begin{tabular}[c]{@{}l@{}}Automatically \\ (Python Program)\end{tabular} &
  \begin{tabular}[c]{@{}l@{}}Automatically \\ (Python Program)\end{tabular} \\ \midrule
\textbf{\begin{tabular}[c]{@{}l@{}}Limitation of \\ reference data \\ generation method\end{tabular}} &
  Costly &
  Costly &
  \begin{tabular}[c]{@{}l@{}}Need for interpretation \\ of topic model outputs\end{tabular} &
  \begin{tabular}[c]{@{}l@{}}Need for interpretation \\ of topic model outputs\end{tabular} &
  Costly \\ \midrule
\textbf{\begin{tabular}[c]{@{}l@{}}Evaluation \\ sample size\end{tabular}} &
  \begin{tabular}[c]{@{}l@{}}125 posts\\ 2,511 comments\end{tabular} &
  \begin{tabular}[c]{@{}l@{}}125 posts\\ 2,511 comments\end{tabular} &
  47,873 concerns &
  47,873 concerns &
  100 concerns \\ \midrule
\textbf{Actual data size} &
  \begin{tabular}[c]{@{}l@{}}65,377 posts\\ 1,392,776 comments\end{tabular} &
  \begin{tabular}[c]{@{}l@{}}65,377 posts\\ 1,392,776 comments\end{tabular} &
  47,873 concerns &
  47,873 concerns &
  \begin{tabular}[c]{@{}l@{}}58,728 concerns \\ (classification)\\ 47,873 concerns \\ (prevalence)\end{tabular} \\ \midrule
\textbf{Results} &
  0.55* &
  0.78* &
  0.80* &
  \begin{tabular}[c]{@{}l@{}}0.95* (k=1)\\ 1.00* (k=2)\end{tabular} &
  \begin{tabular}[c]{@{}l@{}}0.74* (classification)\\ 0.82* (prevalence)\end{tabular} \\ \bottomrule
\end{tabular}%
}
\caption{Metrics Chosen to Evaluate LLM-Powered Analysis Phases and Results. Metric values belong to the closed interval {[}0,1{]}, with higher values indicative of better performance.  * values are statistically significantly different from chance measured via a binomial test (p-value \textless 0.05). 					}
\label{tab:llm-evaluation}
\end{table}

\subsubsection{Rationale for using LLMs:}
We considered using statistical text analysis techniques like LDA, however, they overlook linguistic nuances due to their ``bag of words'' model, failing to capture crucial narrative elements like context in word order. Similarly, BERT derivatives, while better at context capture, are limited to processing 512 tokens at a time. Furthermore, both of these topic modeling methods require meticulous hyperparameter tuning, e.g., choosing number of topics apriori, and significant human effort to translate output topics to human readable themes, making them unsuitable for large datasets. Ultimately, using these established methods will necessitate reliance on multiple disparate components (LDA, human interpretation, text classifiers), leading to complexity, fragility, and evaluation challenges.

LLMs offer a unified and scalable solution to circumvent issues when analyzing relatively unstructured text data, like  online forum posts. Recently published works and preprints like TopicGPT \cite{pham2023topicgpt} and LLooM \cite{lam2024concept} demonstrated the benefits of LLMs over traditional topic models in terms of performance and scalability \footnote{Although these works are similar to our own prompting strategy, they surface topics, rather than themes, which is the focus of our work. We consider ``themes'' to be different from ``topics''. A topic refers to the subject matter or main idea on a surface level described in a word or few phrases, while a theme explores deeper insights and left open to interpretation.}. We believe scalability is crucial for several reasons: it enables prioritizing intervention points based on the frequency of community concerns, is necessary for tasks where breadth is preferred over depth, and captures a diversity of opinions. However, recognizing the limitations of LLMs in nuanced understanding, our novel approach blends the scalability of LLMs with the depth of human expertise and interviews.

\section{Limitations} We recognize the following limitations of our methods:

\noindent \textit{(i) Sample Size:} We conducted interviews with 9 participants, reflecting recruitment challenges for in-person studies with this demographic. Despite the small sample, we reached thematic saturation \cite{caine2016local}. Moreover, the expansive Reddit community analysis complemented our insights into drivers' concerns. Hence, although our analysis may not describe every driver’s experience, the evidence we have collected nonetheless reveals structural features of the Uber system that could potentially affect any driver using the application.

\noindent \textit{(ii) Geographical Location:} Most interviewees were located within a 100-mile radius of the lead author's university in the North-East U.S. Further, we draw solely upon data from English-language forums on Reddit and drivers interviews.
However, the themes identified are applicable across regions to any driver since they concern platform features that are not restricted to specific geographical locations.

\noindent \textit{(iii) Screenshots related to Uber:} Our study exclusively used screenshots from the Uber app for annotation during interviews. However, we emphasize that the underlying algorithmic features are also present in the Lyft app, and most drivers are familiar with both platforms.

\noindent \textit{(iv) Reproducibility Challenges in LLM outputs:} The results obtained from LLMs, like GPT-4, are not guaranteed to be consistent or reproducible, as the model's outputs can vary across different runs, even with the same input data, due to the stochastic nature of the underlying algorithms.

\noindent \textit{(v) Black Box Performance of LLMs:} The performance of LLMs may be poor, necessitating rigorous evaluation methodologies as we have defined. Additionally, LLMs struggle to identify unexpected findings, resulting in a form of collective wisdom. By complementing LLMs with interviews, we can also achieve depth rather than mere breadth.

\section{Positionality Statement}
Our team's expertise spans Human-Computer Interaction (HCI), AI \& Machine Learning, Technology Ethics and Policy. As academics, we advocate for increased transparency in technology platforms. All the authors have engaged with gig economy workers,  policymakers, union organizers and a subset of them are involved in designing decentralized platforms to mitigate AI's negative labor impacts. One of the authors is also a part-time driver with Uber. Our positionality influenced our data interpretation from Reddit and interviews, as well as participants' willingness to engage with us. Acknowledging our unconscious biases and academic position, we exercised reflexivity and adherence to ethical research practices, though some biases inevitably impacted the study.

\section{Ethics Statement on Reddit Data Use}
Reddit users are not representative of the broader population \cite{gaffney2018caveat, jamnik2019use}. Using Reddit data for research purposes introduces risks such as privacy concerns due to exposing sensitive or personally identifiable information\cite{fiesler2024remember}. 
Specific to our work, Reddit participants in rideshare forums may exhibit stronger opinions, be more tech-savvy, or male-dominated \cite{ferrer2021discovering}, differing from the broader rideshare driver population. Moreover, the English-specific discussions on the most popular rideshare subreddits could exclude perspectives from non-English speaking immigrant drivers, a significant rideshare workforce segment\footnote{In 2022, nearly half of all immigrants in the U.S. reported being independent workers\url{https://www.mckinsey.com/featured-insights/sustainable-inclusive-growth/future-of-america/freelance-side-hustles-and-gigs-many-more-americans-have-become-independent-workers}}. While this could affect the generalizability of driver experiences, our analysis still reveals structural insights relevant to any rideshare drivers. 
Finally, we note that the application of AI tools like LLMs to Reddit data can amplify existing biases through their outputs or by propagating them when used as training data.

\section{Findings}
We find that drivers experience a lack of transparency around four key features of rideshare platforms: promotions, fares, routes, and task allocation. For each feature, we detail the harms drivers experience due to the lack of transparency, the mitigation strategies they use, and the design solutions they propose to address transparency issues. A lack of transparency is characterized by missing, misleading, or difficult-to-access information on the platform interface.

We complement our rich qualitative interview data concerning harms with quantitative insights from over 400,000 \texttt{r/uberdrivers} and \texttt{r/lyftdrivers} Reddit community members, across more than 1 million submissions and comments, highlighting the widespread nature of these harms. Specifically, we include a table that presents the most representative transparency harms along with relevant quotes from Reddit, ranked according to their frequency and quantified as a percentage of occurrence across all the 31,340 transparency-related harms: promotions (Table \ref{tab:reddit-promotions}), fares (Table \ref{tab:reddit-fare}), routes (Table \ref{tab:reddit-routing}), and task allocation (Table \ref{tab:reddit-task}). We find that the transparency-related harms across all four codes strongly reinforce the harms surfaced by interview participants.

\subsection{Promotions are presented misleadingly, and drivers want accurate and interpretable information about offers}

\begin{figure}[htb]
\centering
\begin{minipage}{.5\textwidth}
  \centering
  \includegraphics[width=.98\linewidth]{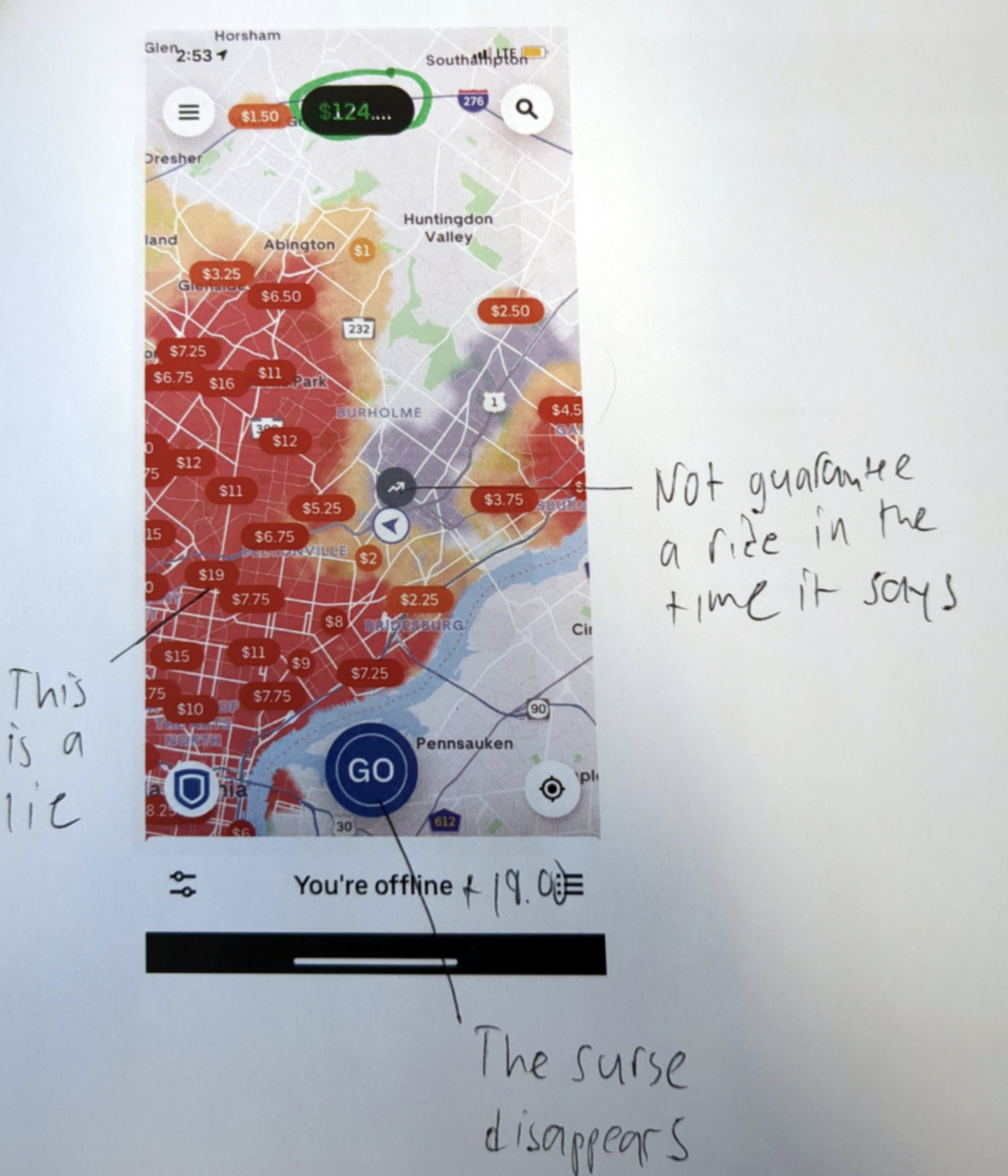}
\end{minipage}%
\begin{minipage}{.5\textwidth}
  \centering
  \includegraphics[width=.98\linewidth]{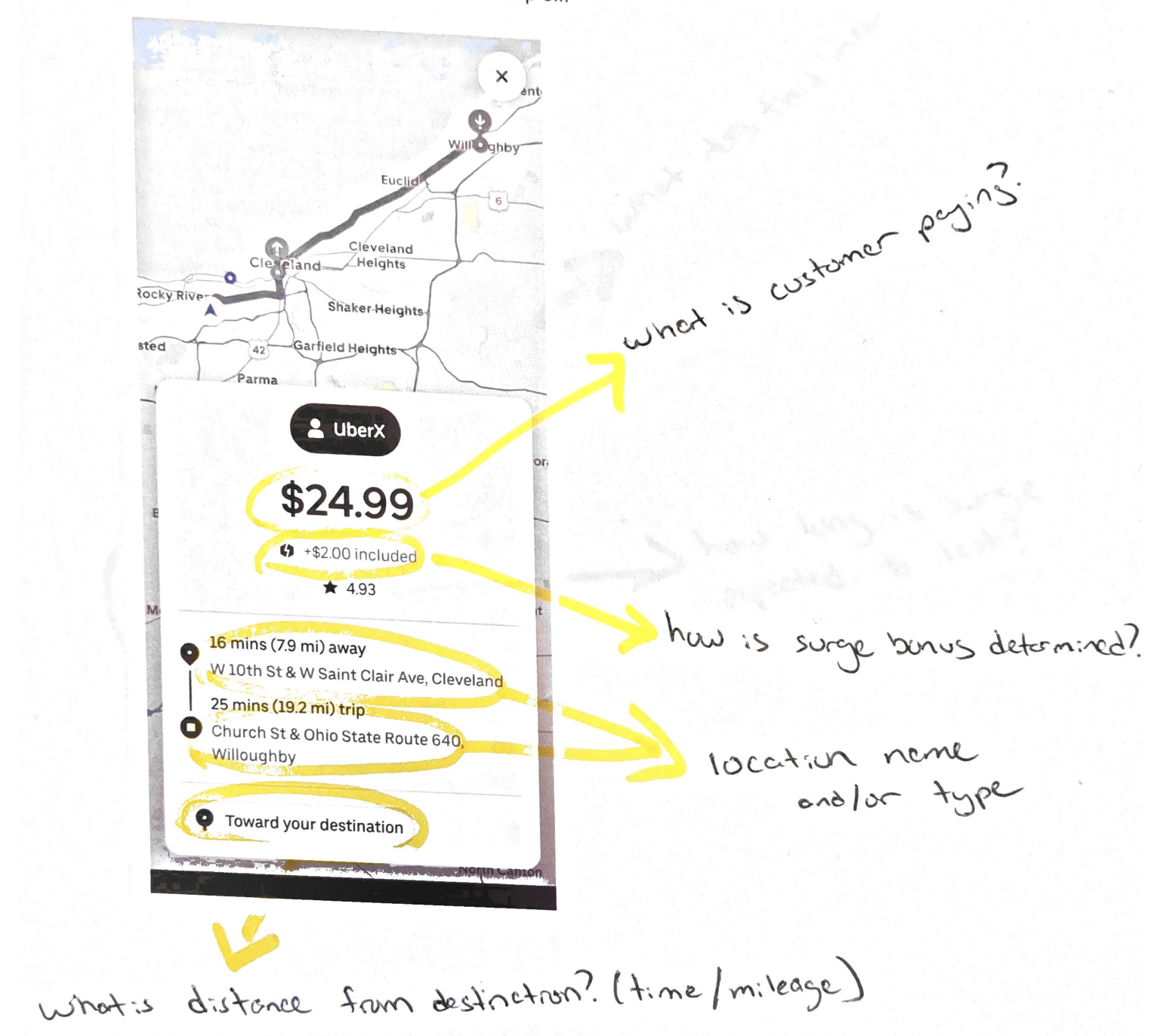}
\end{minipage}
\caption{Annotations of Driver Concerns for Surge Pricing (left) and Ride Offer (right)}
\label{fig:surge_offer}
\end{figure}

As we detail in Section \ref{sec:background}, rideshare platforms offer a variety of gamified promotions, including surges, quests, and boosts. These promotions allow drivers to increase their earnings by meeting a set of criteria. Our participants told us that while these promotions seemed attractive when they initially started driving, over time, they became frustrated with what they characterized as misleading information about promotion offers. 
\subsubsection{\textbf{Harms} caused by lack of transparency in promotions}
\begin{table}[htb]
\small
\begin{tabular}{@{}l|l|c@{}}
\toprule
\multicolumn{1}{c|}{\textbf{Harm}} & \multicolumn{1}{c|}{\textbf{Example Quote}} & \textbf{\% (Count)} \\ \midrule
\begin{tabular}[c]{@{}l@{}}Issues with understanding and qualifying for\\ incentive programs, bonuses, and quests.\end{tabular} &
  \textit{\begin{tabular}[c]{@{}l@{}}Are they really expecting a thumbnail of a map\\ with zero details of where I'd need to be to have \\ a ride qualify for the Quests to be useful?\end{tabular}} &
  \begin{tabular}[c]{@{}c@{}}21\%\\ (6,679)\end{tabular} \\ \midrule
\begin{tabular}[c]{@{}l@{}}Drivers face uncertain and fluctuating income\\ due to changes in surge and quest pricing.\end{tabular} &
  \textit{\begin{tabular}[c]{@{}l@{}}Uber has always done that. When they offer a \\ higher promo, they manipulate and lower surges\end{tabular}} &
  \begin{tabular}[c]{@{}c@{}}9.5\%\\ (2,953)\end{tabular} \\ \midrule
\begin{tabular}[c]{@{}l@{}}Confusion persists over surge pricing\\ determination and boundaries.\end{tabular} &
  \textit{\begin{tabular}[c]{@{}l@{}}I get a ton of riders telling me they are paying \\ $40-$60 for the ride (surging rate), but when the \\ ride ends, I get around \$6 (non-surge rate pay)\end{tabular}} &
  \begin{tabular}[c]{@{}c@{}}6\%\\ (1,858)\end{tabular} \\ \bottomrule
\end{tabular}%
\caption{Most representative transparency harms on Reddit relevant to promotions (Total = 11,490). The harms agree with those surfaced by interview participants.}
\label{tab:reddit-promotions}
\end{table}

Drivers experienced financial and psychological harm due to the lack of transparency in promotions. Surge pricing is presented as a heatmap within the app's interface (see Figure \ref{fig:surge_offer}), where the dark red areas correspond to higher surge multipliers and, thus, higher potential earnings for drivers completing rides within that area. However, there is a \textbf{lack of clarity on boundaries in the surge map} which reduces the utility of the surge map interface as P8 points out: \textit{``So the whole surge map is just terrible because you don't know where you need to be on the map to actually get the money.'' P8}. Uncertainty about information presented on the surge pricing map makes it difficult for drivers to form efficient work strategies and impacts their earnings. 

Drivers face similar issues with misleading information about quests that financially impact them. 
Drivers told us that \textbf{the way information about quests is presented in the app is misleading because it fails to clearly explain how much additional money drivers can earn by completing the bonus}. P3 says:\begin{quote}
    \textit{``So [Uber] can make that more transparent and stop being deceptive...Because [Uber will] tell you, `We'll give you a bonus if you do 40 rides like for \$200.00.' But if you already made your \$200.00 in your 40 rides, you don't get anything.'' P3}
\end{quote}
The way quest amount is currently presented is misleading because drivers do not get the amount \textbf{on top} of their earnings, but rather used to \textbf{top off} earnings if they earn less than that amount for the required amount of rides. However, drivers say this caveat is unclear on the quests offerings page.

Drivers share a common belief that \textbf{gamified promotions were intended to get them to work during times outside of their preferred hours} by offering them attractive offers when they were offline. This perception is in line with how algorithmic nudging works: using soft control as a way to manipulate the supply of drivers~\cite{rosenblat2016algorithmic}. Our participants re-iterated a common finding in the gig work literature–namely that constant nudging and gamification impart a psychological toll on workers~\cite{rosenblat2016algorithmic}. P1 characterized these gamifying techniques as creating a work environment that resembles a casino: \textit{``The house always wins kind of thing. It goes back to the gambling ones, like they're going to give you something that you know is impossible.'' P1}. A lack of clear information about how gamified promotions work and their implications for driver earnings leads to financial and psychological harm.

\subsubsection{\textbf{Mitigation strategies} used to address lack of transparency in promotions}
To cope with unpredictable earnings due to non-transparent gamified promotions, drivers initially experimented with different strategies to identify profitable trip offers. Experienced drivers, like P9, learned to \textbf{avoid chasing surges due to its unpredictability and prefer to accept surges that align with their routine}, explaining, \textit{``[I] absolutely [do] not [chase the surge]...because I know it changes with supply and demand.'' P9} Similarly, drivers like P5 initially tried to maximize earnings by strategically entering surge areas, saying, \textit{``I may go offline and try to get to the \$19 area.'' P5} However, they soon realized the inefficiency of this approach, with P5 noting the frustration of surges diminishing upon arrival: \textit{``I expect that by the time I get there [\$19 surge] will probably be \$8 and then you know you have wasted time.'' P5}. This experience led drivers to avoid promotions, as P2 shares, \textit{``I don't even try to shoot for any higher [value] quest.'' P2}. Similarly, P5 started optimizing for what they perceived to be more reliable trips: \textit{``once I've got a...trip from the airport, I turn on my destination filter from there, and work back towards the airport. And because the airport is a popular destination, it usually works.'' P5}. In general, drivers responded to opaque and uncertain payoffs from gamified promotions by \textbf{developing a preference for predictable and transparent trip offers}.

\subsubsection{\textbf{Design solutions} to address lack of transparency in promotions}
Drivers expressed a desire for design solutions that addressed the lack of transparency in promotions by asking for \textbf{more explanations about how promotions worked}. P3 illustrates this through their experience with the quest feature: \textit{``So they can make [the bonus amount] more transparent... I mean, they do tell you on the second page that, but that's not right, though.'' P3}. Here, P3 calls out how information on how much drivers can expect to earn from a bonus is hidden. They argue that platforms should explain that the \textbf{bonus amount is already factored in the displayed upfront fare, not in addition to it}. As a result, P3 asked that this caveat be shown upfront with the offer. P6 similarly calls for information about \textit{``how many rides make up a boost?'' P6} so drivers better understand how they can increase their earnings through participating in these incentive offers. 

Drivers also suggest that the platform could message them how much they can make from a surge to get a better picture of their potential earnings: \textit{``Before the surge they can send SMS to drivers [saying] `This is what potentially you can make.' I think that... will motivate a lot of drivers as well. Because people always want to gain information, they can make extra money.'' P7}. 

In summary, the platform interface could more clearly show that amounts earned from quests and boosts are supplemental, not additional, to driver earnings, and they could provide \textbf{more granular information about the requirements to qualify for a promotion}. 

\subsection{Fare information is difficult to decipher and drivers want more accurate and explicit metrics}
Drivers earn a portion of the trip fare for each trip they complete. However, the total trip fare and the amount they can expect to earn are often unclear to drivers. The trip fare information shown to drivers depends on whether they work in a region that uses ``upfront pricing'' or ``rate card pricing''. 
Drivers who worked in regions with upfront pricing are shown estimated total earnings as well as some information about a trip's duration and destination. Drivers who worked in rate card regions are shown information about trip duration and distance, and they have to mentally calculate their earnings based on the rate per minute and rate per mile for their region. 
Regardless of the pricing model, drivers told us that the fare breakdowns they saw were inaccurate and made it difficult to accurately predict their earnings. This causes financial harm and also raises questions among drivers about fairness. 

\subsubsection{Harms caused by lack of transparency in fare information}

\begin{table}[htb]
\resizebox{\textwidth}{!}{%
\small
\begin{tabular}{@{}l|l|c@{}}
\toprule
\multicolumn{1}{c|}{\textbf{Harm}}                        & \multicolumn{1}{c|}{\textbf{Example Quote}}                                            & \textbf{\% (Count)} \\ \midrule
Drivers are unsure about opaque fare calculation methods. & \textit{\$30 to the middle of nowhere and had to eat dead miles to drive back} & 23 (7,202)           \\ \midrule
\begin{tabular}[c]{@{}l@{}}Drivers express discontent with the percentage of fares they\\  receive compared to what the platform charges customers.\end{tabular} &
  \textit{\begin{tabular}[c]{@{}l@{}}Yeah it's always high ride demand and they charge the passengers high\\  prices and drivers earnings are like 25\%\end{tabular}} &
  \begin{tabular}[c]{@{}c@{}}3\\ (1,025)\end{tabular} \\ \midrule
\begin{tabular}[c]{@{}l@{}}Drivers are not fairly paid for extensive wait times during \\ stops or added tasks\end{tabular} &
  \textit{\begin{tabular}[c]{@{}l@{}}The stop system is bullshit. You make less money, 100\%. I just checked. \\ To order two rides, one there, one back, \$17. To order the same ride, but\\  using stops it's \$13.\end{tabular}} &
  \begin{tabular}[c]{@{}c@{}}3\\ (856)\end{tabular} \\ \bottomrule
\end{tabular}%
}
\caption{Most representative transparency harms on Reddit relevant to fare information (Total = 9,083). The harms agree with those surfaced by interview participants.}
\label{tab:reddit-fare}
\end{table}

Lack of transparency in fare information makes it difficult for drivers to predict their earnings and work towards financial security. P8, who works in a rate card region, told us that it is \textbf{difficult to decipher exactly how much they could expect to earn based on the fare breakdowns they are given}:\begin{quote}
    \textit{``If you can do the math in your head by looking at the miles and the time, you can possibly figure [the fare] out, but you only have a few seconds really to choose whether you want to accept the ride... A lot of people, myself included, can't do those calculations.'' P8} \end{quote}
P8 describes a situation echoed by other drivers: the \textbf{information presented combined with the available time for processing was insufficient to make an informed decision about potential earnings from a trip offer}. Drivers working in upfront pricing regions do not have to do these mental calculations as they are given an estimate of how much they will earn from a trip. 
However, the upfront price often doesn't account for real-time ride conditions such as wait times, stops, and traffic, as P5 told us: \textit{``I only got paid like \$13, but I spent a whole hour in traffic.'' P5.} P5 summarizes the frustration that even though upfront pricing tells drivers their estimated earnings in the trip offer, these estimations do not necessarily result in financial security or greater predictability with earnings because they \textbf{don't take into account real-time driving conditions} and thus can be misleading. Drivers are also frustrated with high and unpredictable service fees, typically not visible to them, P7: \textit{``You're supposed to take 10\%. But because of all additional fees, they take more than 30\%. Which is a problem for a lot of drivers.'' P7.}

\subsubsection{Mitigation strategies used to address lack of transparency in fares}
The lack of transparency in fares makes it difficult for drivers to calculate and predict their earnings accurately. Drivers confront and try to mitigate the effects of unpredictable earnings with a few strategies that, while resourceful, remain significantly limited without substantive platform-led changes. Full-time drivers are forced to \textbf{work extended hours}, as P4 says \textit{``But you know what...you gotta force yourself to go and do'' P4}, to offset pay disparities. P9 echoes this sentiment by stating that, \textit{``the algorithm...keeps track of how much money you're making. At the end of the day you notice that I'm not making more than 20 bucks an hour... as a full-time driver, you better put in those hours, a lot of hours. So the flexibility right, that is gone.'' P9.} Drivers employ strategies such as \textbf{switching apps or relocating} to find better fares. P9 describes using both Uber and Lyft: \textit{``I have my other app, Lyft, you know. I turn [it] on and let it give me a ride. You turn off Uber.'' P9.} P4 mentions moving to a different area for better offers: \textit{``OK, let me drive, like, 10 miles away from there and then turn back on the app.'' P4.} These strategies, while showcasing drivers' adaptability, underscore the need for platform-level reforms to address unpredictable earnings genuinely.

\subsubsection{\textbf{Design solutions} to address lack of transparency in fares}
Drivers want to be able to predict their earnings from a trip offer. Rate card pricing put a cognitive load on drivers as they had to do quick mental calculations to estimate how much they would make. These drivers emphasized the importance of being shown your earnings upfront. 
The upfront model gives drivers the transparency they desire and reduces the cognitive load on drivers to calculate fares mentally. However, this model is not used consistently across the US, and drivers point out that the estimated earnings can be misleading. As a result, drivers proposed that rideshare companies could \textbf{show estimated earnings upfront and trip metrics such as distance, time, rate per mile, and rate per minute after a ride} has occurred so they could assess if their earnings felt fair, \textit{``I think you can have upfront pricing and have a mileage rate at the same time. I mean, they kind of already do it with the reservations.'' P5.} P4 suggested a model like the one taxi cab companies use where: \textit{``they started off with the base pay. And then there was a certain per mile you went, and then there was also a charge for per minute that you were inside that vehicle.'' P4}. These fare breakdowns would make it easier for drivers to forecast their earnings and understand how fares were calculated.

However, showing rate breakdowns in more detail does not necessarily address the core issue of financial precarity that drivers face, which is one of their key concerns. Instead, \textbf{drivers want easily interpretable, straightforward pricing models where trip metrics, like mileage and time, correspond closely to their earnings}. P4 summarizes this sentiment, saying: \textit{``If it was fair pay, honestly, then, a lot of these issues wouldn't be a thing. You know, if the mileage and the time equated to the pay that you deserve without any additional bumps or anything like that, then I don't think there will really be much of an issue.'' P4}. 
These statements underline a shared desire for a straightforward and fair pay structure. 
Ultimately, the inadequate pay coupled with the lack of trust in the platforms underscore the critical role of transparency for drivers, considering their challenging labor conditions.%

\subsection{Route information is imprecise and drivers want accurate, contextualized, and reliable geographic information}
When considering trip offers, drivers typically have fewer than 10 seconds to accept or decline a ride. In that limited time, they have to parse and decide based on the trip metrics presented in the platform interface (see Figure \ref{fig:surge_offer} for an example of a trip offer). Drivers point to issues with the accuracy and completeness of geographic information presented and the amount of time given to accept or decline an offer. Drivers told us that the lack of transparency in trip offer information leads them to engage in unsafe driving behaviors and negatively impacts their financial well-being.
\subsubsection{\textbf{Harms} caused by lack of transparency in route information}

\begin{table}[htb]
\resizebox{\textwidth}{!}{%
\small
\begin{tabular}{@{}l|l|c@{}}
\toprule
\multicolumn{1}{c|}{\textbf{Harm}} &
  \multicolumn{1}{c|}{\textbf{Example Quote}} &
  \textbf{\% (Count)} \\ \midrule
\begin{tabular}[c]{@{}l@{}}Concerns over inadequate payment for time and \\ distance invested in long pickups and waits.\end{tabular} &
  \textit{\begin{tabular}[c]{@{}l@{}}I don’t understand what the hell is Uber thinking when giving\\  us long pickups for short trips. NOBODY sane accepts a \\ \$4 ride for someone half an hour away.\end{tabular}} &
  \begin{tabular}[c]{@{}c@{}}5\\ (1,429)\end{tabular} \\ \midrule
\begin{tabular}[c]{@{}l@{}}Drivers experience technical difficulties with app \\ navigation and route mapping, leading to disruptions.\end{tabular} &
  \textit{\begin{tabular}[c]{@{}l@{}}I started using the Lyft nav, then Google maps, then Waze only\\ to go back to the Lyft nav. I went through that cycle because the\\ Lyft nav sucks but at least you see everything going on with the app. \\ With the others, I’ve lost out on streaks and pings because of glitches, etc.\end{tabular}} &
  \begin{tabular}[c]{@{}c@{}}4\\ (1,208)\end{tabular} \\ \midrule
\begin{tabular}[c]{@{}l@{}}Drivers express risks due to dangerous multitasking\\  and unsafe areas suggested by algorithms\end{tabular} &
  \textit{\begin{tabular}[c]{@{}l@{}}If a Pax wants to be taken to a part of town I don’t feel safe going, \\ I feel like I should know, for my own safety and peace of mind.\end{tabular}} &
  \begin{tabular}[c]{@{}c@{}}3\\ (909)\end{tabular} \\ \bottomrule
\end{tabular}%
}
\caption{Most representative transparency harms on Reddit relevant routing information (Total = 3,546). The harms agree with those surfaced by interview participants.}
\label{tab:reddit-routing}
\end{table}

Drivers tell us that getting trip offers is distracting, but they need to pay attention to their phones while they drive if they want to earn money. P2 describes this tension, saying: \textit{``I'm literally on the bridge, like going 60 or whatever... and they literally send me a request, like what do I do?'' P2}. 
Drivers utilize indicators such as trip distance, estimated duration, and end pickup/dropoff location to decide whether to accept a trip offer. However, drivers claim that these indicators are often inaccurate, not present, or incomplete. For example, multiple drivers told us that they are \textbf{unable to see the full address for the pickup and dropoff location}, which makes it difficult for them to assess if a trip is worthwhile, \textit{``it's hard to really know if this is the right town that you're actually going in, because in New Jersey it's so many towns, it could be two towns with the same name'' P3}. Drivers voiced \textbf{concern for their physical safety when driving without knowing beforehand where they were going and said they felt compelled to decide between safety and making a living}: \textit{``If you want to stay out of a bad neighborhood...you will never make any money doing this job.'' P1}. P2 explained their frustration about the lack of transparency with trip destinations, saying: \textit{``If you're not a high enough tier, you get less information [about the trip]. So you don't even know exactly, like, where the trip is going.'' P2}. Making \textbf{access to precise geographical features of a trip contingent on a driver's tier status}, exacerbates issues of transparency because a driver's access to information is contingent on a convoluted algorithmic process that determines their tier status. 

Even though rideshare platforms have features such as \textbf{``destination filters''} that purport to help drivers by enabling them to filter for trip offers around a destination the driver selects, drivers told us that these features \textbf{rarely work as intended}. P4, P5, and P6 criticize the malfunctioning destination filter, as P6 states: \textit{``Every time I use it, it sends me away from my house. I live close to Philadelphia, it may try to send me to New York.'' P6}. 
  
\subsubsection{\textbf{Mitigation strategies} used to address lack of transparency in route information}
Drivers utilize a variety of strategies, including \textbf{auto-accepting all ride offers} and compromising on physical safety to cope with inadequate information processing time and incomplete geographic details. P9 explains their approach to handling incoming trip requests and inadequate information processing time by saying: \textit{``What happens if you lose a [trip] request? Your acceptance rate gets hit. Because of that, I set up the app for automatically accepting the ride.'' P9}. This strategy enables them to stay focused on the road and not worry about impacts on their acceptance rate if they miss a trip offer while driving. However, this meant that sometimes P9 accepted trips that paid very little for the amount they had to drive. Several drivers mentioned they just ask the passenger where they're going and follow their suggestion rather than relying on the app, P6: \textit{``I cater to the people, to be honest with you...I depend on the actual passengers a lot of times, so I ask them [where they're going]'' P6}.

Ultimately, drivers navigate a precarious balance between ensuring their safety and optimizing for earnings. Some, like P6, will \textbf{not wait long in unsafe neighborhoods}: \textit{``If I go to a bad neighborhood. You know, I don't get a response, you know, within a couple of minutes, then I'm definitely trying to pull off.'' P6} And P7 discusses the importance of understanding passenger psychology for self-protection. Yet, drivers like P1 often feel \textbf{compelled to accept rides in risky areas}, as declining can affect their earnings, and the algorithm may offer the same ride again: \textit{``I usually roll the dice and take the person... like 3:00am in the morning at the bottom ... which is like a really bad neighborhood, you gotta decide if it's worth it because you're gambling ... that person might try to rob you.'' P1} However, P1 is quick to add that \textit{``You don't have a choice; otherwise you will make no money. So the safety is like you just have to get over it.'' P1}.

\subsubsection{Design solutions to address lack of transparency in route information}

\begin{figure}[htb]
\centering
\begin{minipage}{.5\textwidth}
  \centering
  \includegraphics[width=.98\linewidth]{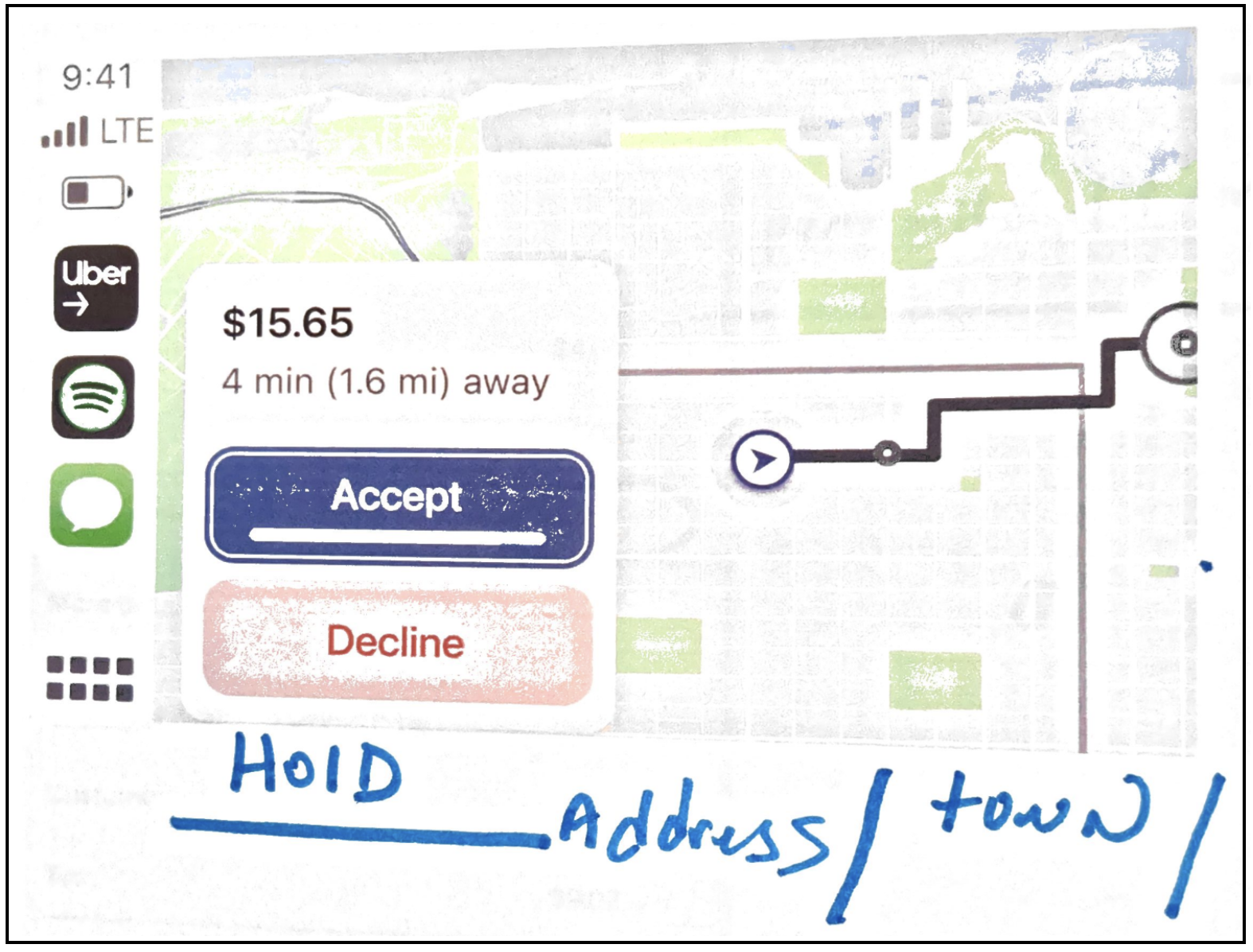}
\end{minipage}%
\begin{minipage}{.5\textwidth}
  \centering
  \includegraphics[width=.98\linewidth]{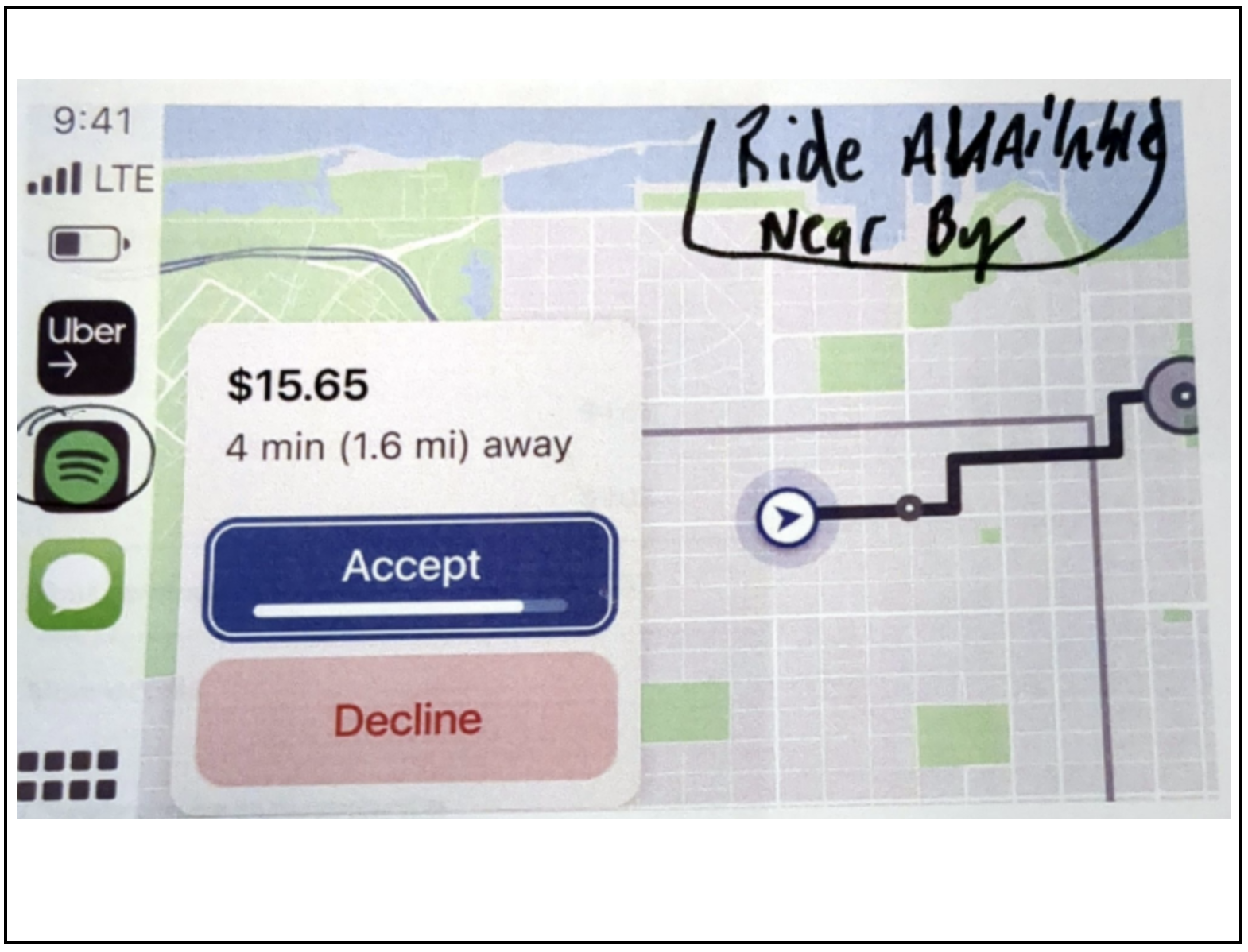}
\end{minipage}
\caption{Hold button (left, P3) and Ride Available Nearby feature (right, P6). P3 and P6 propose these interventions to prioritize safety over algorithmic management during a ride.}
\label{fig:safety_buttons}
\end{figure}

Drivers express a need for more time to make informed decisions about ride offers. P4, P6, P7, and P9 also stress that their focus should be on the road during a ride, not on making calculations or decisions. P4 specifically notes, \textit{``really my focus needs to be on the road, not trying to do math in my head and be looking at my phone!'' P4}. As a result, P3 and P6 propose the \textbf{implementation of a Pause/Hold Button and a complementary Ride Available feature}, as seen in Figure \ref{fig:safety_buttons}. P6 describes a system where a ride offer pops up only when the driver has stopped, enhancing safety: \textit{``If it pops up a ride available, when you stop, for safety reasons, you can look at the ride and say I can do it or I can't do it.'' P6} P3 further elaborates on this idea, suggesting a hold or pause button that allows drivers to pause offers during a ride or a break (e.g., coffee/restroom breaks) and review those offers once they have safely stopped: \textit{``Maybe we have like a hold button or a pause button or something that you can click on... Put in the hold until you safely stop ... I think this is what's causing a lot of the accidents. So many people are looking at the phone. Let me put it on hold, and when I stop, if it's still available, I could take a look at it or what they have available in the area.'' P3}.

Drivers call for platforms to \textbf{add more contextualized geographic information}, create features that let drivers safely review and accept ride offers, and consistently display a set of basic information for all trip offers. P3 draws on their experience asking passengers for landmarks near their destination: \textit{``But what I try to do is I ask the customer...which school are you going to or, you know which town...give me an idea.'' P3} to inform their design solution to include contextual geographic like the town's name or landmarks in the proximity of the dropoff location, helps the driver better understand where they're going. 

\subsection{\textbf{Task allocation} is black-boxed and drivers want explanations about how they work}
Due to the opaque nature of rideshare platforms' AI and algorithmic decisions, drivers lack insight into the factors influencing the quality and quantity of trip offers. While drivers believe acceptance/cancellation rates, ratings, and reviews impact trip quality, the lack of transparency in task allocation prevents definitive proof. This opacity results in psychological and financial harm, hindering drivers' ability to establish a predictable work routine. It also raises fairness concerns, as drivers are uncertain about the criteria determining task quality. To address this, drivers seek design features displaying how their in-app metrics contribute to task allocation, aiming for transparency and improved understanding.

\subsubsection{Harms caused by a lack of transparency in task allocation}

\begin{table}[htb]
\resizebox{\textwidth}{!}{%
\small
\begin{tabular}{@{}l|l|c@{}}
\toprule
\multicolumn{1}{c|}{\textbf{Harm}} &
  \multicolumn{1}{c|}{\textbf{Example Quote}} &
  \textbf{\% (Count)} \\ \midrule
\begin{tabular}[c]{@{}l@{}}Drivers express concerns over clarity in ride \\ assignments and earnings determination by algorithms.\end{tabular} &
  \textit{\begin{tabular}[c]{@{}l@{}}A driver who keeps a high AR, low cancel rate, gets mostly 5 Stars, \\ and gives out mostly 5 Stars to pax will get better (more lucrative) rides\\  than a driver the company's algorithm doesn't like.\end{tabular}} &
  \begin{tabular}[c]{@{}c@{}}9\\ (2,765)\end{tabular} \\ \midrule
\begin{tabular}[c]{@{}l@{}}Challenges with understanding and adapting to algorithm\\ changes that affect ride assignments and income.\end{tabular} &
  \textit{\begin{tabular}[c]{@{}l@{}}I multi-app, but the algorithm has changed. If you log off \\ from the app, to ride for another app, the algorithm puts you \\ on the bottom of the queue after you log on again for the next ride.\end{tabular}} &
  \begin{tabular}[c]{@{}c@{}}8\\ (2,408)\end{tabular} \\ \midrule
\begin{tabular}[c]{@{}l@{}}Concerns over lack of clarity in cancellation criteria\\ and the associated fare impact.\end{tabular} &
  \textit{\begin{tabular}[c]{@{}l@{}}Then I ask about the cancellation fee and they said they \\ can’t provide one as no option came up.\end{tabular}} &
  \begin{tabular}[c]{@{}c@{}}4\\ (1,197)\end{tabular} \\ \midrule
\begin{tabular}[c]{@{}l@{}}Uncertainty about how ratings are determined and their\\ direct impact on drivers' work opportunities.\end{tabular} &
  \textit{\begin{tabular}[c]{@{}l@{}}I read that Lyft will try to match a pax with a driver that both had \\ 5 starred each other in the past.\end{tabular}} &
  \begin{tabular}[c]{@{}c@{}}1\\ (384)\end{tabular} \\ \midrule
\begin{tabular}[c]{@{}l@{}}Drivers feel platform algorithms unfairly influence ride\\ distribution, surge pricing, and overall income.\end{tabular} &
  \textit{\begin{tabular}[c]{@{}l@{}}A lady was standing next to my car and ordered a lyft and it gave her\\ a driver that was outside of the Bonus zone and 20 minutes away.\end{tabular}} &
  \begin{tabular}[c]{@{}c@{}}1\\ (316)\end{tabular} \\ \midrule
\begin{tabular}[c]{@{}l@{}}Concern over algorithms distributing rides unfairly\\ or prioritizing certain drivers\end{tabular} &
  \textit{\begin{tabular}[c]{@{}l@{}}Perceived favoritism by the app's algorithm towards newer drivers\\ can limit earnings for long-term drivers.\end{tabular}} &
  \begin{tabular}[c]{@{}c@{}}0.5\\ (151)\end{tabular} \\ \bottomrule
\end{tabular}%
}
\caption{Most representative transparency harms on Reddit relevant to task allocation (Total = 7,221). The harms agree with those surfaced by interview participants.}
\label{tab:reddit-task}
\end{table}

Drivers accepted and completed trips they would have otherwise deemed unsafe because they were \textbf{afraid of impacting their acceptance/cancellation rates} and being punished by the platform's algorithm. For example, P5 told us that, ``as a woman,'' she was hesitant to accept trips in certain neighborhoods late at night, but she was \textit{``not going to cancel the trip for the sake of not trying to damage my rating'' P5}. Drivers worry that their performance ratings are affected by their demographic characteristics. For example, P7, a Black man, said: \textit{``Some people... they act certain kind of way based on their race, based on their background, based on where they live... some drivers get a lot of bad ratings.'' P7}. He further wondered whether these \textbf{ratings and reviews may indirectly influence the algorithm's matching and compensation outcomes}. 

\begin{figure}[htb]
\centering
\begin{minipage}{.5\textwidth}
  \centering
  \includegraphics[width=.98\linewidth]{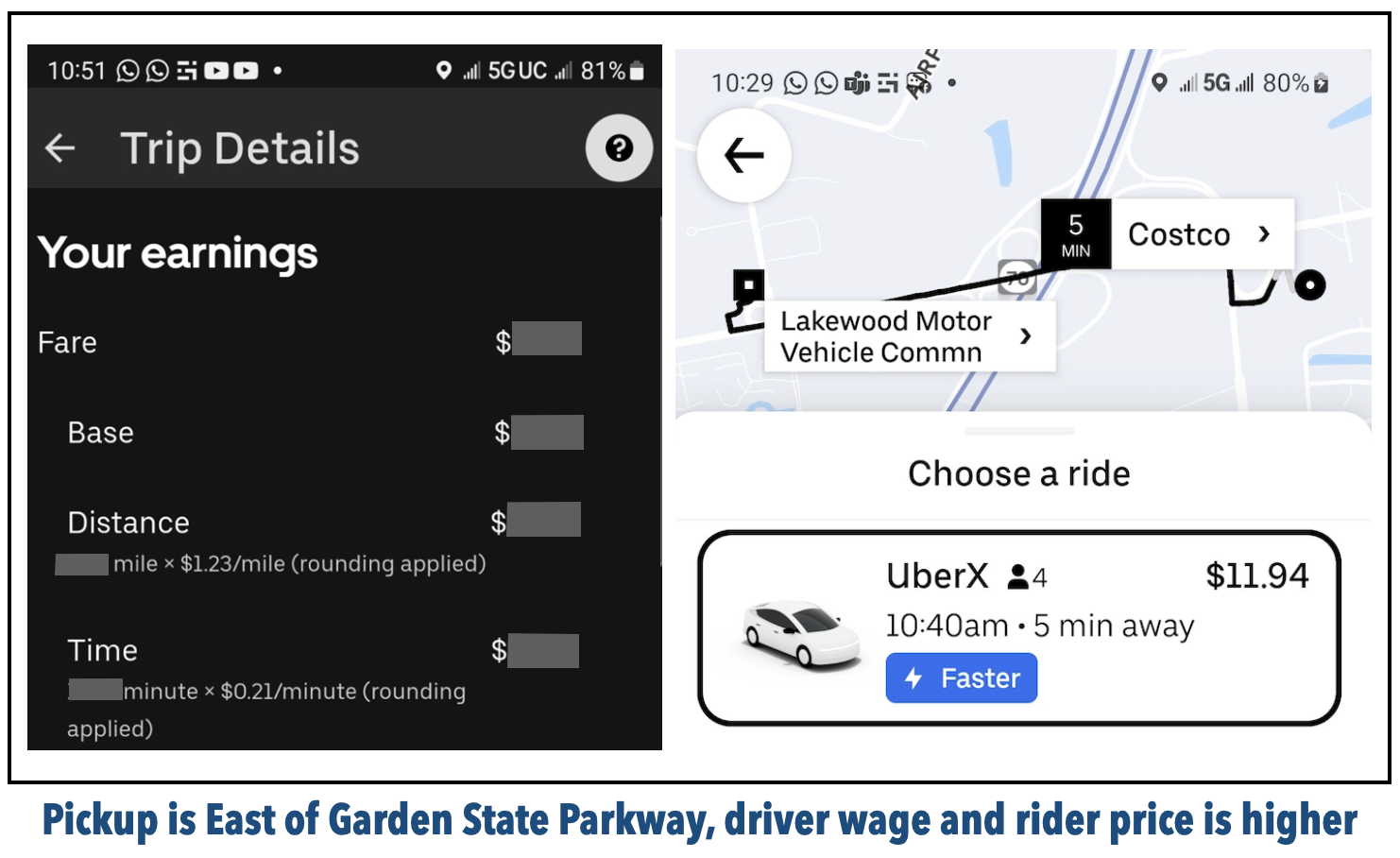}
\end{minipage}%
\begin{minipage}{.5\textwidth}
  \centering
  \includegraphics[width=.98\linewidth]{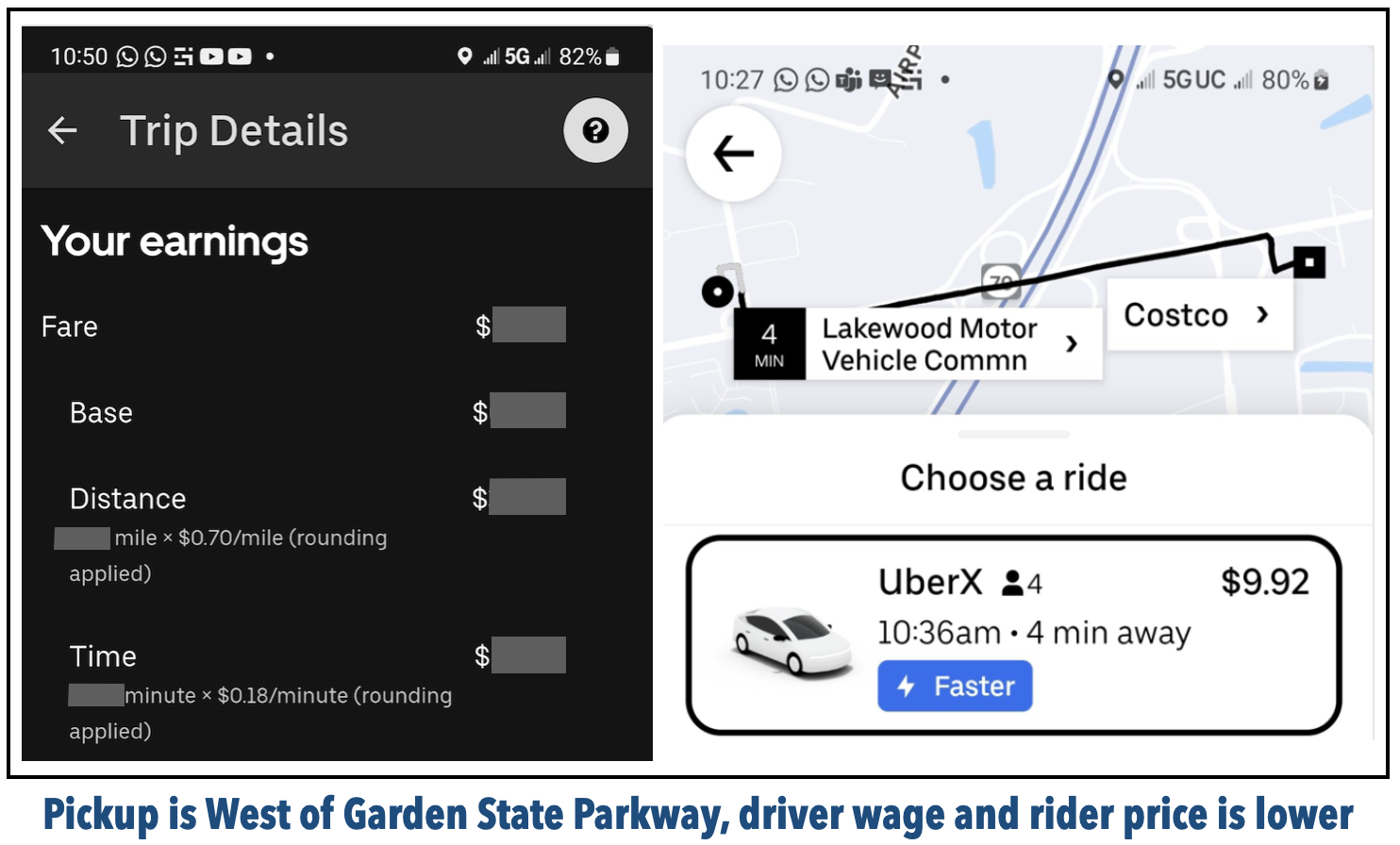}
\end{minipage}
\caption{Evidence of algorithmic wage and price discrimination shared by P8. The earnings are redacted to preserve participant privacy.}
\label{fig:algo_wage_discrimination}
\end{figure}

Without insight into how task allocation works, drivers worry about potential discrimination and unfairness. P2 told us how they believed \textbf{algorithmic wage discrimination–different pay for similar work–was taking place}, saying: \textit{``if you had two different phones and you got two different drivers, one person was getting one thing and somebody else was getting another ... and I was like, that can't be right ...it was the same distance, just different people ... why is one person offered a higher rate and why is another person offered a lower rate?'' P2} P8 shared screenshots demonstrating evidence of algorithmic wage discrimination. Figure \ref{fig:algo_wage_discrimination} shows different rate cards for similar rides across the Garden State Parkway in NJ. The rate card showed \$1.23/mile and \$0.21/minute for pickups from the east, but only \$0.70/mile and \$0.18/minute from the west, even though the pickup spots may be less than a mile apart. P8 noted, \textit{``From Lakewood MVC West of the Parkway to Costco [on the] East, Uber charges \$9.92 one way, and \$11.94 the other." P8} This evidence aligns with other drivers' experiences and also highlights the algorithm's role in wage (and price) discrimination.

\subsubsection{Mitigation strategies used to address lack of transparency in task allocation}
Drivers choose to \textbf{optimize their in-app metrics (e.g., acceptance rates and performance rating)} in hopes of remaining in good standing with the platform and thus getting more and better tasks. Some drivers, like P9 use the auto-accept feature: \textit{``What happens if you lose a request? Your acceptance rate gets hit. Because of that, I set up the app for automatically accepting the ride.'' P9}. Others, like P3, go to lengths to be polite with unruly passengers out of fear that the passenger could give them a low rating and impact their performance rating. Some drivers, like P2 and P8, attempt to scrutinize the payouts for tasks by comparing trip offers with each other. In summary, drivers attempt to mitigate opacity in task allocation by maximizing the two in-app metrics most visible to them, acceptance rate and performance rating, and through \textbf{systematic collective examination of trip offers}.

\subsubsection{Design solutions to address lack of transparency in task allocation}
Drivers suggest ways that rideshare platforms could use in-app metrics for task allocation in a way that allows drivers to exert agency without fear of reprisal. Drivers emphasize that their \textbf{cancellation rate should not be impacted by cancellations made for safety reasons during a ride}. P7 insists that such cancellations, made while driving, should not affect their cancellation metrics: \textit{``Once you're driving already, if you make cancellations, it should not affect your cancellation [rate].'' P7.} P5 extends this argument to cancellations due to low passenger ratings, advocating that safety considerations should override concerns about metrics: \textit{``I don't think we should be docked or penalized or should go towards our cancellation rate for cancellations due to passengers that have low ratings and cancellations that we do while we're driving.'' P5}. 

Drivers also asked for \textbf{explanations about how the acceptance and cancellation rates impacted the trips they were offered} as they felt that having lower acceptance/cancellation rates would lead to less desirable trip offers, but they struggled to gather empirical evidence that would disprove this hypothesis: \textit{``like what [is Uber] basing that off of? I feel like since I started driving a little bit more full time, I haven't gotten [a quest bonus] in quite some time.'' P4}. P4 illustrates their uncertainty when trying to understand what led to promotions. P2 also told us that they \textbf{wanted explanations about how reward amounts for promotions are calculated} so they could make decisions about how often they needed to drive: \textit{``I mean how the rewards are being calculated? But these numbers change a lot, so I have no idea how these rewards are calculated.'' P2}. 
In summary, drivers want explanations of how task allocation decisions are made so they can make informed decisions about their work strategies. They want explanations to be provided post-hoc to enable adequate time for reflection.

\section{Lessons for Regulating Rideshare Platforms: Rideshare Platforms Must Publish Transparency Reports}
\label{sec:transparency}

\begin{figure}[htb]
    \centering
    \includegraphics[width=\columnwidth]{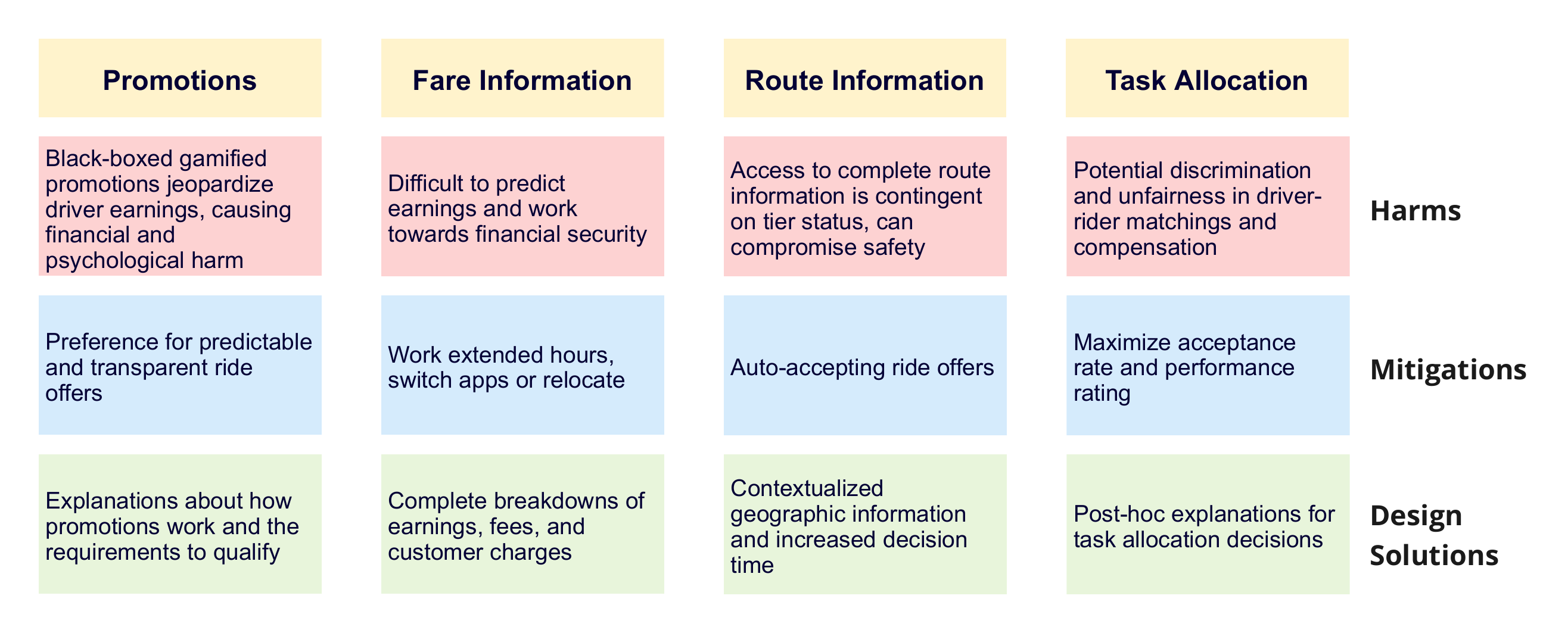}
    \caption{Thematic analysis map: Overview of transparency-related findings from driver interviews and Reddit data analysis.}
    \label{fig:findings-interview-overview}
\end{figure}

We present an overview of our findings in Figure \ref{fig:findings-interview-overview}, which highlights a transparency gap between existing platform
designs and the information drivers need, particularly concerning promotions, fares, routes, and task allocations.

\subsection{Design Solutions Require Organizational Consideration and Policy Intervention}

Drivers need transparency into algorithmic decisions, such as promotions, fares, routing, and task allocation. While rideshare drivers propose design changes for transparency, rideshare platforms, driven by a focus on scale and maximizing profits, may resist implementing them without proper incentives. Moreover, information asymmetries are beneficial to rideshare platforms as they enable platforms to extract surplus value from each transaction by having access to information unavailable to other parties due to the platforms' position as marketplace designers~\cite{viljoenDesignChoicesMechanism2021}. This disincentives platforms to balance information asymmetry through increased transparency.

While our participants identified innovative design features that could increase transparency in their work, it is important to consider how rideshare platforms' operational logics constrain these design solutions. We build on recent work that argues that researchers must consider organizational contexts, actively incorporate policy, and conduct the necessary translation work to firmly ground policy recommendations in empirical research findings and put forth a \textit{transparency report} geared toward policymakers exploring the impact of AI on labor~\cite{yang2024future, dalal2023understanding}. This report demonstrates how technology researchers can engage in translation work between empirical studies and policymaking to address the misalignment between platform interests and drivers' transparency needs.

\subsection{Rideshare Transparency Reports as a Solution to Drive Regulatory Changes}
Transparency has long been essential for clarity and social accountability \cite{kalderon2015form, florini2007right, acemoglu2013nations}, with its importance recognized in various fields \cite{johnston2006good, schudson2015rise}. Its growing relevance in digital technologies, especially in social media and AI \cite{Heikkilä2023high, diresta2022time, bommasani2023foundation, edelson2021standard, silverman2023tiktok}, is leading to calls for transparency through legislative actions like the EU's Digital Services Act and AI Act, and the US's proposed Platform Accountability and Transparency Act. 

Transparency is a powerful catalyst for change. Transparency initiatives in social media catalyzed legal actions and ad practice reforms. Facebook's Ad Library\footnote{Facebook Ad Library: \url{https://www.facebook.com/ads/library/}} enabled research \cite{nagaraj2023discrimination, le2022audit, papakyriakopoulos2022algorithms} and lawsuits\cite{trucks2022fb}, resulting in settlements which have changed targeted job and housing ad systems\cite{vrs2022fb}. Pay transparency narrowed gender pay gaps\cite{pay2022transparency}. These precedents show transparency drives material change. 

For rideshare platforms, enhanced transparency can also boost predictability, fairness, and safety, driving regulatory changes. Mandating platform data transparency enables research, fair wage advocacy, treating platforms as utilities, and policies addressing low gig pay. However, currently, rideshare platforms currently lack incentives to voluntarily publish the granular data necessary to enhance transparency, absent legislation requiring disclosure \footnote{Chicago's rideshare regulations mandate platform data disclosures, establishing a policy precedent; see \url{https://www.chicago.gov/city/en/depts/bacp/provdrs/vehic/svcs/tnp.html}, and \url{https://chicago.github.io/tnp-reporting-manual/}}. 

Building on existing policy precedents in AI, social media regulations, and nascent rideshare regulations, we argue researchers must pursue diligent translation between technical design interventions and policy domains. The transparency report would help realize a subset of the technical interventions we have proposed by providing access to data that only platforms currently control. We propose a concrete outline and establish baseline reporting guidelines for a robust rideshare transparency report. These findings highlight critical implications for policymakers. We call for rideshare platforms to publish comprehensive \textit{transparency reports} detailing indicators for workers, advocates, and regulators to better understand the complex rideshare landscape and make more informed decisions that benefit drivers. 

\subsection{Platform Inclusion Criteria}
A rideshare platform should publish transparency reports if it meets any of the following four criteria:

\begin{enumerate}
    \item Algorithmically\footnote{We use the term algorithmically to mean the use of an AI-based or other algorithmic system} determines and varies the wages of drivers for similar rides defined by distance and/or duration
    \item Algorithmically matches drivers and riders
    \item Algorithmically varies the platform fee for each ride
    \item Algorithmically incentivizes drivers through gamification features (e.g., quests, challenges, bonuses), which vary across drivers according to, but not limited to, location, tenure, and demand.
\end{enumerate}
    
Uber and Lyft are examples of platforms that meet all of the above-mentioned criteria.

\subsection{Transparency Report Indicators}
We define 51 indicators that characterize transparency in rideshare platforms inspired by our analysis of Reddit and interview data. We divide the indicators into 4 broad categories: indicators about \textit{ride statistics}, indicators about \textit{driver statistics}, indicators about \textit{algorithmic management}, and finally indicators about \textit{platform policies}. We provide an overview of the indicators here and a comprehensive list in the Appendix Figure \ref{fig:transparency-report}. 
\begin{enumerate}
    \item \textbf{Ride Statistics (24 indicators):} These indicators include timestamps of pickup, dropoff, waits and stops, overall ride duration and distance, and a breakdown of fees and wages. These ride statistics indicators are vital for enabling algorithmic audits and enhancing overall transparency, specifically regarding the breakdown of wages, to investigate questions of algorithmic wage discrimination. 
    \item \textbf{Driver Statistics (11 indicators):} These indicators include drivers' earnings, ratings, reviews, acceptance and cancellation rates, and demographics. This should be available for all service locations. These indicators are crucial for enhancing predictability, bolstering driver agency, and promoting wage parity across the platform.
    \item \textbf{Algorithmic Management (9 indicators):} These indicators include inputs, outputs, and data used in the platform's algorithms, privacy mitigations, incentives (quests, surges), and impact assessments on drivers, enhancing and realizing driver needs of explainability and predictability.
    \item \textbf{Platform Policies (7 indicators):} These indicators include details about safety incidents, customer service inquiries, new platform features, maximum and minimum earnings, rate cards, as well as deactivations and appeals, providing essential transparency into the rideshare ecosystem, extending beyond direct AI and algorithmic decisions.
\end{enumerate}

\subsection{Data Access and Retention} For effective regulation and transparency, the FTC and local transportation authorities (e.g., see data on Transportation Network Providers in the Chicago Data Portal \footnote{The City of Chicago Transportation Network Providers - Trips (2018 -2022): \url{https://data.cityofchicago.org/Transportation/Transportation-Network-Providers-Trips-2018-2022-/m6dm-c72p/about_data}}), should either obtain data directly from rideshare platforms or mandate these platforms to publish and maintain it in a centralized data warehouse. This warehouse should regularly update, providing daily, weekly, monthly, and yearly data snapshots and summaries. It must offer a publicly accessible API, enabling requests by date range, platform, and location, and a web interface for basic data searches. Platforms are advised to retain data for one year, while government agencies should maintain it for seven years, balancing the need for error correction and data integrity against storage costs.

\subsection{Recommendations for Policymakers} As policymakers shape the future of rideshare platforms, a focused approach to transparency, enforcement, and realistic policy goals is crucial. We provide 3 recommendations to policymakers below. Ultimately, these recommendations call for a more accountable and sustainable transportation sector.

\begin{enumerate}
    \item \textbf{Prioritizing Transparency in New Rideshare Regulation:} Transparency should be a foremost consideration in rideshare regulation and should mandate the publishing of rideshare transparency reports. This prioritization will enhance accountability and public trust in these services.
    \item \textbf{Enforcing Existing Transparency Policies:} Policymakers should actively enforce current legislation through data protection authorities and the FTC \cite{zanfir2019dpa, ftc2022gig, hess2019transparency} to increase transparency in rideshare platforms. This enforcement can lead to more informed decisions by platforms, drivers, riders, and regulators and foster better practices within the industry.
    \item \textbf{Setting Realistic Expectations for Transparency} While transparency remains vital, policymakers must acknowledge its constraints. Worker data collectives, third-party applications, and application design changes represent potential complementary solutions to improve driver well-being. Therefore, policies should strive for a reasonable transparency level, recognizing it as one facet of a broader strategy addressing rideshare challenges.
\end{enumerate}

\subsection{Transparency is Not a Panacea: Limitations and Unintended Consequences}
Researchers have highlighted significant limitations of transparency initiatives across domains; we acknowledge them here. \citet{keller2021humility, urman2023transparent, husovec2024digital} note the lack of standardization in transparency data disclosures, allowing companies discretion over measurement approaches without adhering to established principles (e.g., Santa Clara Principles). Further, companies lack incentives to devote resources for dedicated transparency infrastructure, leading to insufficient disclosure \cite{dubois2021micro, keller2021humility}. This can result in ``transparency washing''\cite{zalnieriute2021transparency}, where initiatives obfuscate substantive questions about corporate power. Encapsulating all these points, \citet{ananny2018seeing} develop a 10-point typology examining transparency's limitations in creating accountable algorithmic systems. Their typology argues revealing system parts and data alone is insufficient for understanding or governance. Instead, engaging with the limits of transparency can help us use it as a way to hold sociotechnical systems accountable by focusing on the relationships and contexts involved, rather than just making things visible.

Mandating transparency could also inadvertently undermine core business interests and result in unintended consequences\cite{seattle2024uturn}. Disclosing competitive algorithms like surge pricing models could diminish platforms' competitive advantages. Furthermore, transparency requirements may force pricing models to converge, reducing competition and consumer choice. Publicly sharing driver incentives could open avenues for collusion and market distortions. Revealing surge zones may result in oversupply in those areas, diminishing earning potential elsewhere. Privacy risks can also emerge, without adequate safeguards, from disclosing driver and rider location data. Taken together, transparency initiatives face inherent trade-offs between accountability and adverse effects, and policymakers need to balance the needs of stakeholders carefully.

\section{Conclusion and Future Work}

Our study exposes a lack of transparency in rideshare platform features, leading to financial, emotional, and physical harm to drivers. Through a unique mixed-methods study that combines in-person interviews with nine drivers and an LLM-based analysis of over one million Reddit posts, our findings highlight the lack of transparency in key rideshare platform features, including promotions, fares, routes, and task allocation. Importantly, the coherence between the interview and Reddit data underscores the widespread and persistent nature of transparency-related harms within the rideshare community. The significance of our study lies in not only uncovering transparency harms, mitigation strategies, and needs of rideshare workers, but also proposing concrete solutions. Specifically, we advocate for policymakers to call on rideshare platforms to regularly publish rideshare transparency reports, echoing successful precedents in social media and AI regulation. These reports, informed by workers' transparency needs surfaced through our study, detailing crucial indicators such as ride statistics and algorithmic insights, can empower drivers and address their transparency needs. Taken together, this study contributes to the ongoing discourse on platform regulation and design, moving beyond discussions of information asymmetry to envisioning platform design where transparency is prioritized, empowering drivers to address power imbalances and resulting in improved rideshare worker well-being.

Future work can investigate jurisdictions with established rideshare platform regulation (e.g., Germany, Spain, UK \cite{uber2023status}) or active resistance (e.g., Colombia\cite{uber2020colombia}, Brazil\cite{uber2015brazil}) to understand workers' transparency needs. Do transparency demands persist when drivers are classified as employees receiving wages and benefits? Comparative studies across regulatory environments could reveal valuable global insights into the evolving rideshare platform-worker-policymaker dynamics.

\section*{Acknowledgments}
We acknowledge the Azure Cloud Computing Grant from CSML at Princeton which enabled access to Microsoft Azure compute resources. We are grateful to Prof. Lindsey Cameron, Prof. Amy X. Zhang, Prof. Min Kyung Lee, Prof. Toby Jia-Jun Li, Prof. Christian Sandvig, Angie Zhang and Hope Schroeder for their broad discussions. We thank Prof. Arvind Narayanan and Mihir Kshirsagar for their insights on the generalization of this work and its policy implications. We thank Prof. Brian Keegan for pointing us to the Reddit data source. We thank Sayash Kapoor for insightful discussions on the conceptual and normative implications of LLMs and prompting strategies. We are grateful to Vinayshekhar Bannihatti Kumar for his valuable inputs on defining evaluation metrics. We thank Sunnie Kim for her contributions on connections to Explainable AI. We appreciate Elizabeth Anne Watkins for providing valuable critiques from an interpretivist perspective, which helped reframe our work. We thank Yuhan Liu, Beza Desta, and Kiyosu Maeda from the Princeton HCI group for their annotations. We thank the participants and organizers of the CHI LLM as Research Tools 2024 Workshop for their feedback. Finally, we thank the anonymous CSCW and FAccT reviewers for their valuable feedback.

\bibliographystyle{ACM-Reference-Format}
\bibliography{references}
\clearpage

\appendix
\section{Interview Protocol and Screenshots}
\label{asec:protocol-screenshots}
\subsection{Interview Protocol}

\noindent \textbf{1) Introduction}

\noindent Hi everyone, and thank you for coming. Today, we're studying how clear or ``transparent'' rideshare platforms’ behind-the-scenes rules, called ``algorithms,'' are to you. These algorithms use AI and Machine Learning techniques, which are basically computer programs that learn from yours and other drivers driving patterns. We want to understand what you need to know to do your job and any challenges you face when these rules aren't clear. Our ultimate goal is to advocate for greater transparency of these algorithms for your benefit. Today’s activity will be structured based on the 3 stages of a ride – at or before pickup, during the ride, at or after drop off.

\noindent \textbf{2) Ice Breaker}

\begin{itemize}
    \item Take 1 minute to draw a doodle of your self portrait. Write your name, preferred pronouns, how long you’ve been driving for, and what app(s) you use.
    \item We’ll then have you introduce yourself with your self portrait and describe the most interesting ride you’ve ever done, in about 2-3 minutes.
\end{itemize}

\noindent \textbf{3) Interview and Design Activity: Annotate Screenshots and Ask Probing Questions}

\noindent Here are 10 screenshots we found on public forums about various aspects of a ride. We would like you to annotate these images. We’ll progress through the screenshots sequentially.

\noindent This screenshot is about [insert topic e.g., surge]. 
\begin{itemize}
    \item Use the marker to circle information that you felt uncertain about. For example, you might feel uncertain about something because you don't know WHY it's being presented or HOW it was calculated.
    \item Use the pen to write/draw what additional information you would have liked in that context
\end{itemize}

\noindent [For each screenshot]
\begin{itemize}
    \item This screenshot is about [insert algorithmic decision / factor e.g. surge pricing, acceptance rate, etc]. Reflecting on this screenshot and the information you circled in red, could you share why you were uncertain about that piece of information within the context of [algorithmic decision/factor]? 
    \item Can you describe a ride where you directly experienced or observed [insert algorithmic decision] on your app? How did it guide or influence your actions? How did it make you feel? (OR) Reflect on a time when you noticed [insert algorithmic factor] on your app screen. How did this factor influence or shape your decisions?
    \item Reflecting on the information you’ve written using the pen, could you share what additional information or context about [insert algorithmic decision / algorithmic factor] on the screenshot would you have liked?
\end{itemize}

\noindent \textbf{4) Conclusion}

\noindent Alright, before we conclude our interview, I'd like to touch on a few more points based on your annotated screenshots and the discussions we've had regarding the three stages of a ride
\begin{itemize}
    \item What do you do when you feel like you don't have enough information about a ride? Do you turn to third-party apps (Mystro, Maximo, Gridwise, SurgeApp), WhatsApp message groups, or Reddit forums, for example?
    \item As you gained more experience with rideshare work, has your emotional response to not getting adequate information about rides changed? Tell me about the biggest differences between you as a driver when you started vs. where you are today.
\end{itemize}

\subsection{Screenshots and Associated Probing Questions}

\begin{figure}[h]
    \centering
    \begin{minipage}[c]{.5\textwidth}
        \centering
        \includegraphics[width=.9\linewidth]{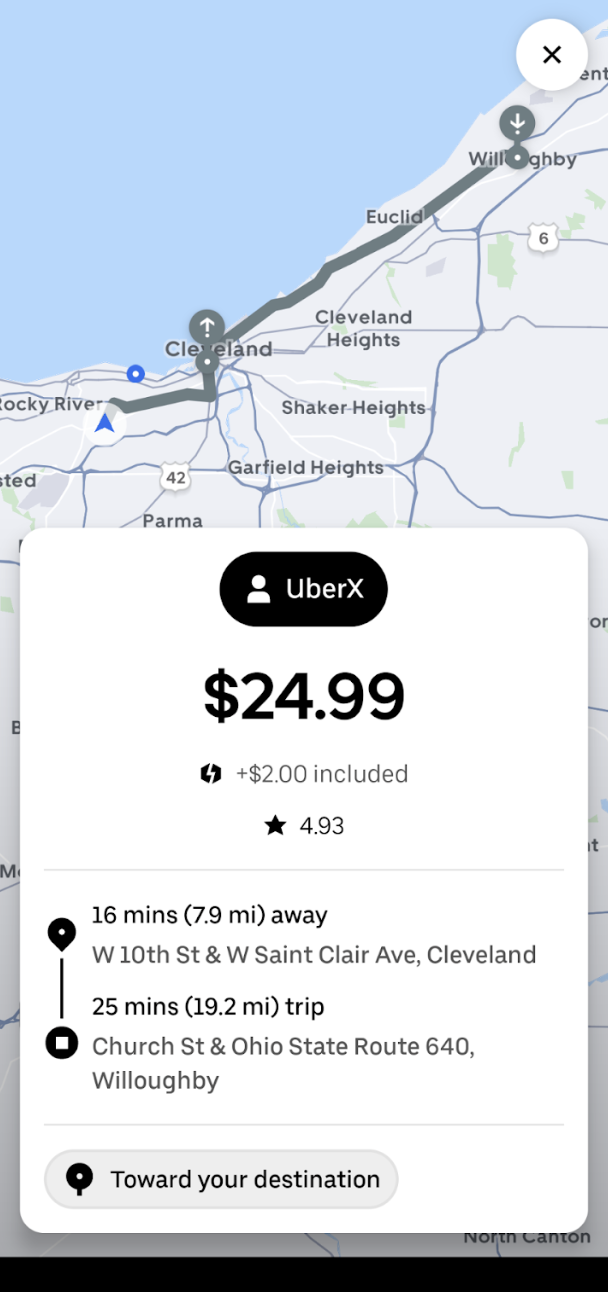} %
        \caption{Screenshot 1}
        \label{fig:screenshot1}
    \end{minipage}%
    \begin{minipage}[c]{.5\textwidth}
        \vspace*{\fill} %
        This screenshot is about the algorithms which determine the upfront price, providing you with an offer and the route to a destination or drop off. \\ \\
        \noindent \textbf{Probing questions} \\
        \begin{itemize}
            \item When you start your work day, how do you decide which app you want to use? What additional information would be useful for you to have when choosing which app(s) to turn on? 
            \item Tell me about how you decide if a trip is worthwhile. What additional information do you wish you had?
            \item Can you tell me about a trip you were offered recently that you thought was really good? What made it ``really good''?
            \item Tell me about how you decide on what route to take for a ride. Can you tell me more about what concerns you have when you need to make adjustments to the suggested route, route closures, traffic etc?
            \item What factors do you consider when you choose the location where you wait for your first ride to come in?  How does this compare to the information provided by the app?
            \item Can you tell me about what it's like to wait for a passenger to show up for a ride? What additional information do you wish you had? How long do you usually wait for?
        \end{itemize}
        \vspace*{\fill} %
    \end{minipage}
\end{figure}

\begin{figure}[h]
    \centering
    \begin{minipage}[c]{.5\textwidth}
        \centering
        \includegraphics[width=\linewidth]{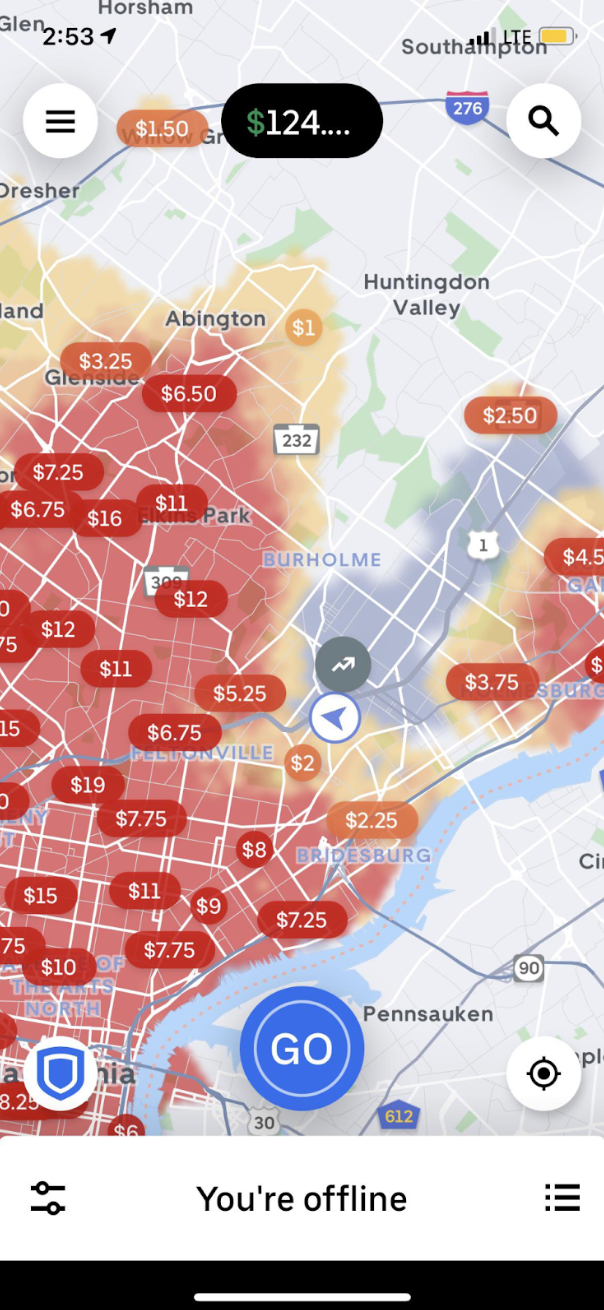} %
        \caption{Screenshot 2}
        \label{fig:screenshot2}
    \end{minipage}%
    \hspace{0.5cm}
    \begin{minipage}[c]{.4\textwidth}
        \vspace*{\fill} %
        This screenshot is about the surge pricing algorithm. \\ \\

        \textbf{Probing Questions} \\ %
        \begin{itemize}
            \item Do you ever consider surge pricing when picking a spot to wait for more requests? 
            \item Can you tell me a little more about how you understand and work with surge?
            \item Where and when do surges usually occur?
        \end{itemize}
        \vspace*{\fill} %
    \end{minipage}
\end{figure}

\begin{figure}[h]
    \centering
    \begin{minipage}[c]{.5\textwidth}
        \centering
        \includegraphics[width=.9\linewidth]{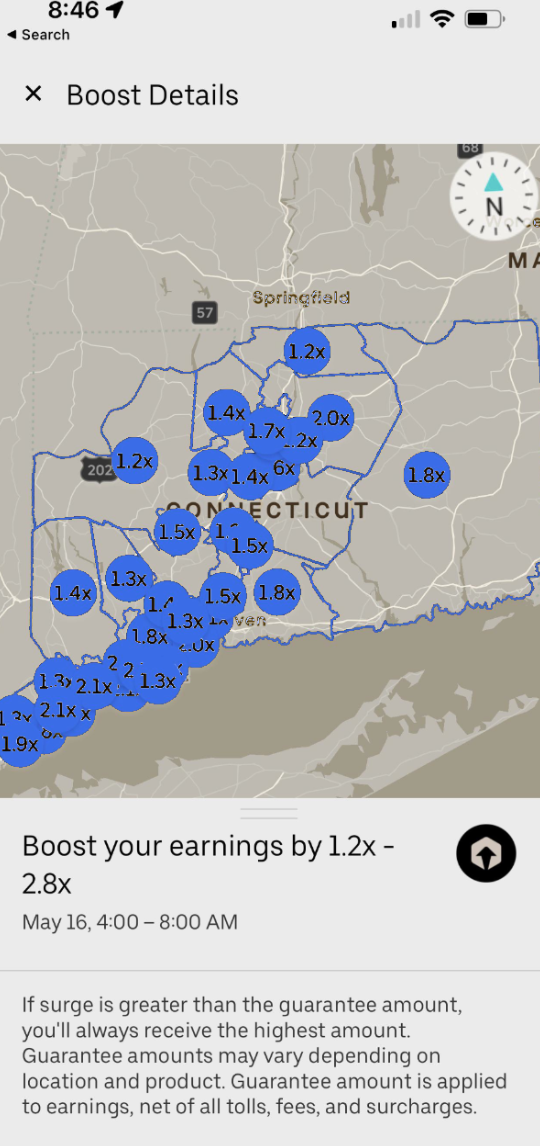} %
        \caption{Screenshot 3}
        \label{fig:screenshot3}
    \end{minipage}%
    \hspace{0.5cm}
    \begin{minipage}[c]{.4\textwidth}
        \vspace*{\fill} %
        This screenshot is about the boost algorithm.\\ \\ 
        \textbf{Probing Questions} \\ %
        \begin{itemize}
            \item Do you ever consider boost pricing when picking a spot to wait for more requests? 
            \item Can you tell me a little more about how you understand and work with boost?
        \end{itemize}
        \vspace*{\fill} %
    \end{minipage}
\end{figure}

\begin{figure}[h]
    
    \includegraphics[width=\linewidth]{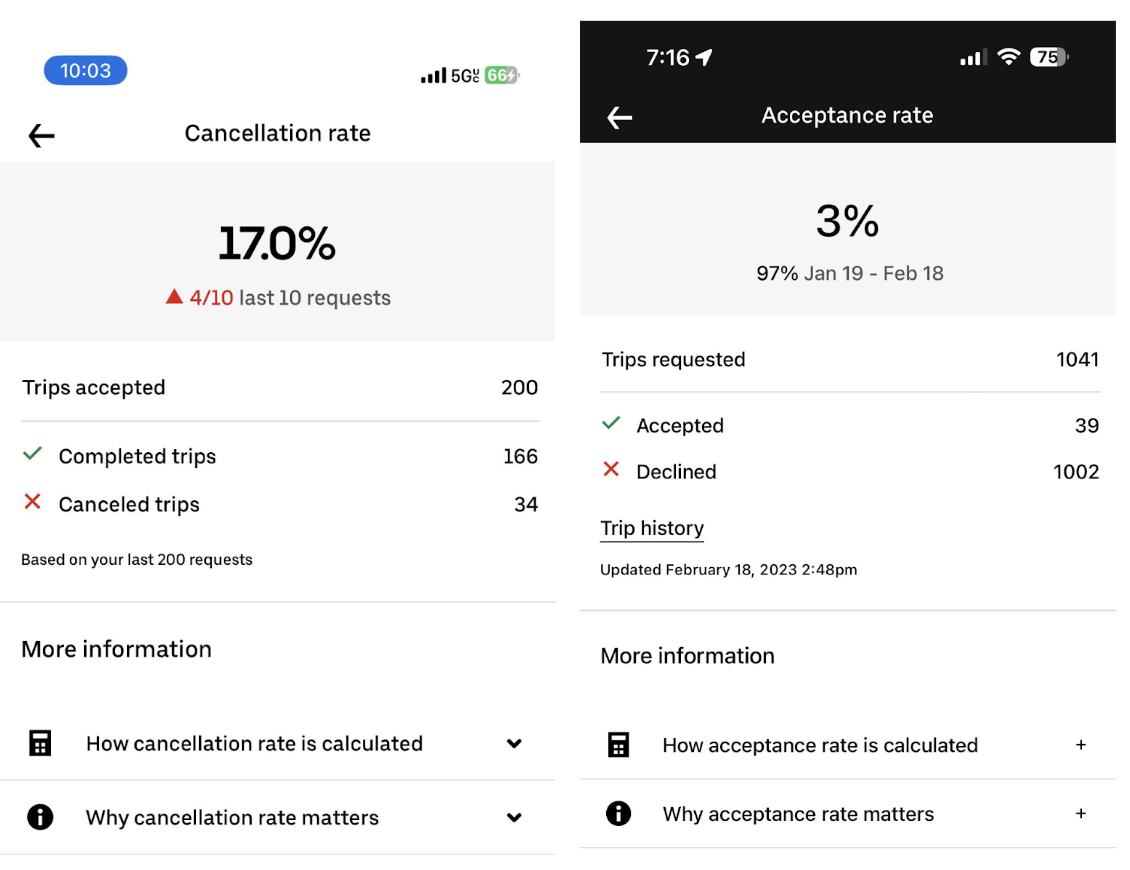} %
    \caption{Screenshot 4}
    \label{fig:screenshot4}  

    \bigskip %

    \begin{minipage}{\textwidth} %
    This screenshot is about the cancellation and acceptance rate.\\ \\
        \textbf{Probing Questions} \\
        \begin{itemize}
            \item How do you think about the acceptance and cancellation rate? 
            \item How do the acceptance and cancellation rates influence your driving habits?
        \end{itemize}
    \end{minipage}
\end{figure}

\begin{figure}[h]
    
    \includegraphics[width=\linewidth]{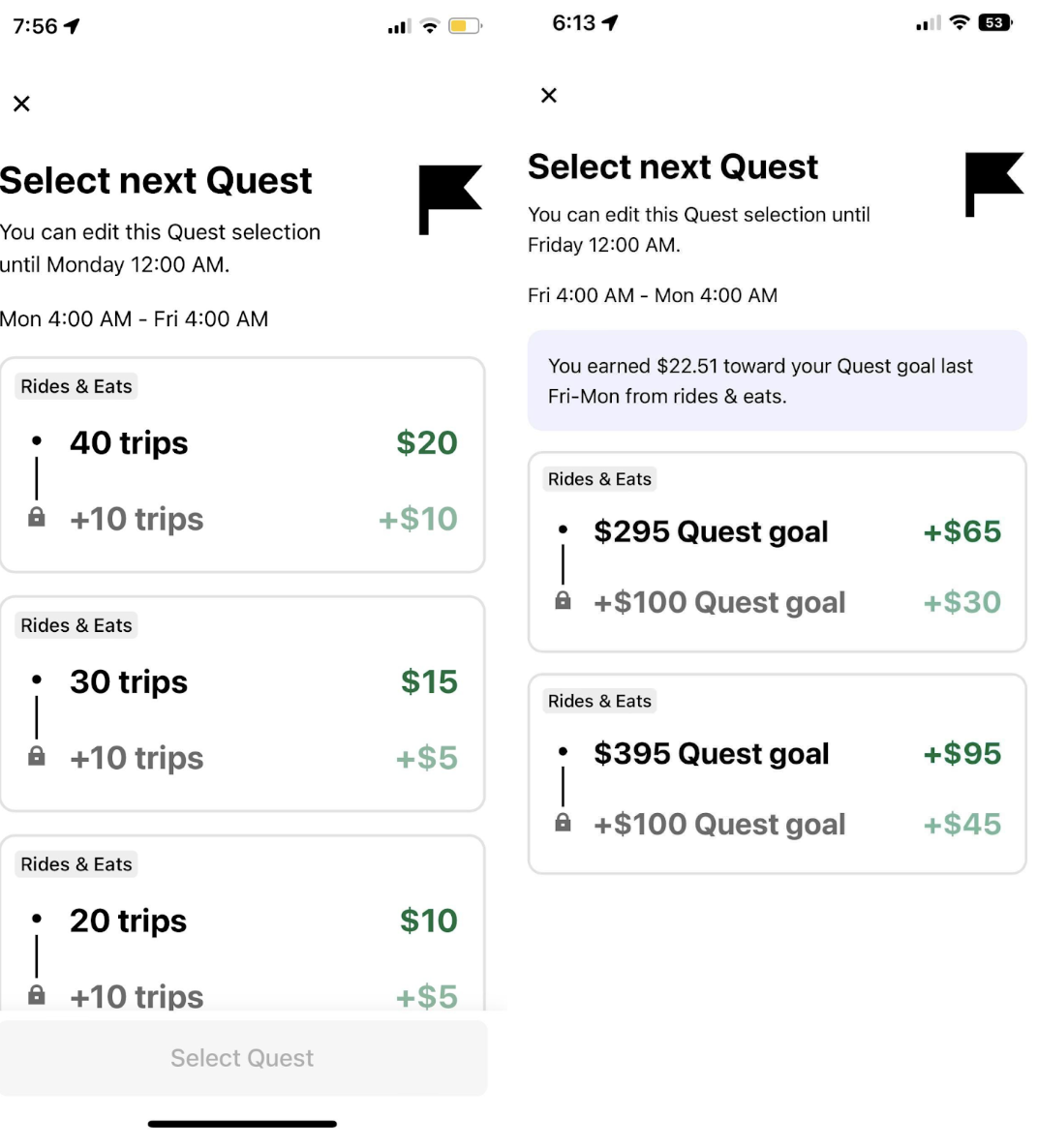} %
    \caption{Screenshot 5}
    \label{fig:screenshot5}  

    \bigskip %

    \begin{minipage}{\textwidth} %
    This screenshot is about quests. \\ 

        \textbf{Probing Questions} \\
        Can you tell me a little more about how you understand and work with quests? How does it influence your selection of trips?
    \end{minipage}
\end{figure}

\begin{figure}[h]
    
    \includegraphics[width=\linewidth]{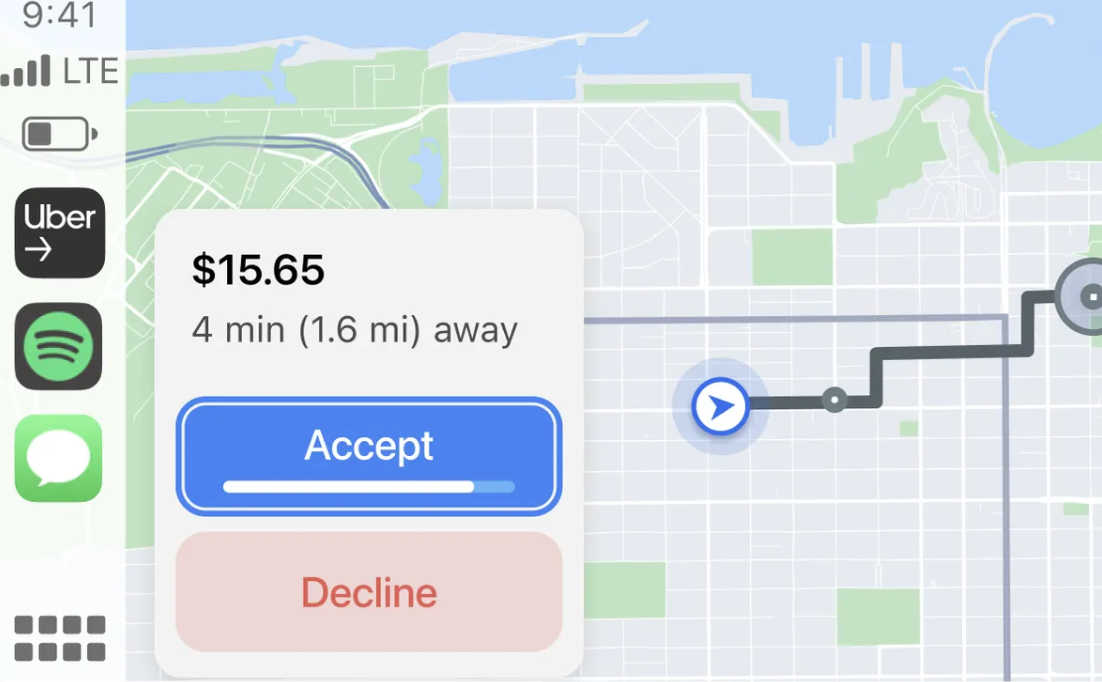} %
    \caption{Screenshot 6}
    \label{fig:screenshot6}  

    \bigskip %
    
    \begin{minipage}{\textwidth} %
    This screenshot if about ride offers you receive while you are on a ride \\ 

        \textbf{Probing Questions} \\
        What is your decision-making process like when you are offered a new ride while driving?
    \end{minipage}
\end{figure}

\begin{figure}[h]
    \centering
    \begin{minipage}[c]{.5\textwidth}
        \centering
        \includegraphics[width=.9\linewidth]{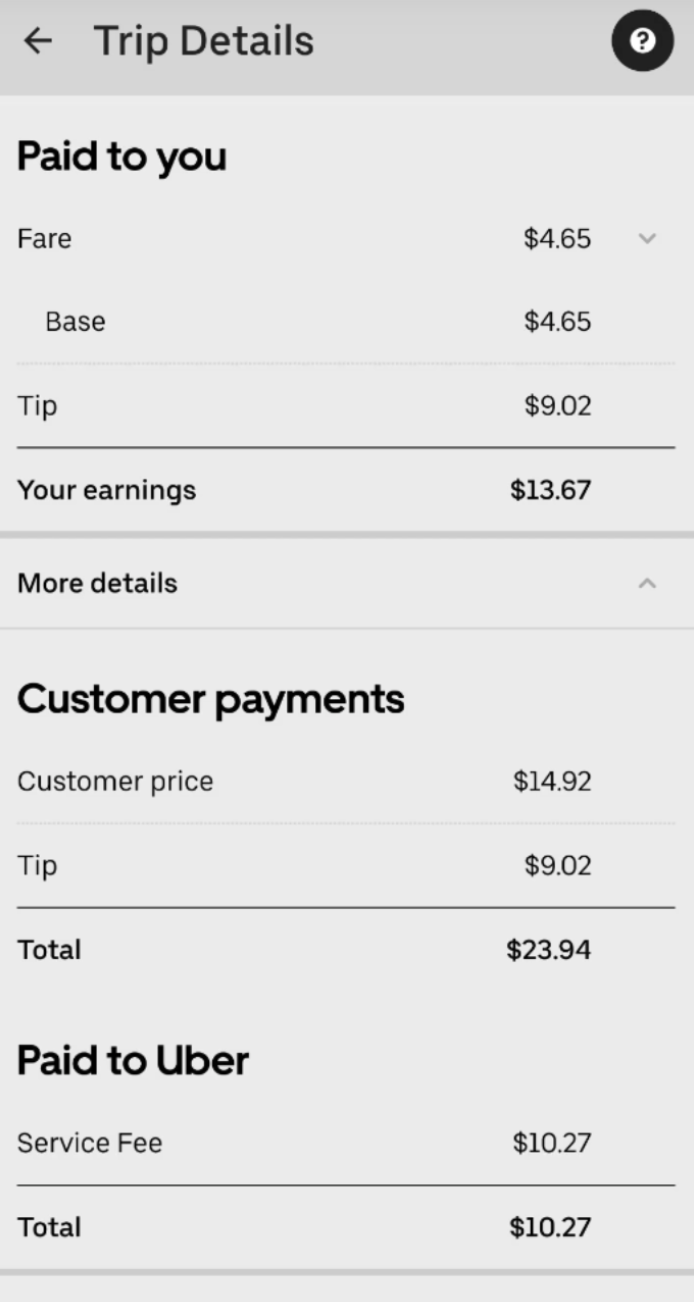} %
        \caption{Screenshot 7}
        \label{fig:screenshot7}
    \end{minipage}%
    \hspace{0.5cm}
    \begin{minipage}[c]{.4\textwidth}        
        \vspace*{\fill} %
        This screenshot is about fare breakdown at the end of a ride.\\ \\
        \textbf{Probing Questions} \\ %
        \begin{itemize}
            \item Can you tell me about how you think about your pay when you’re (1) waiting for a rider before a pickup? (2) stopped for a rider in between a trip?
            \item Have you noticed insurance deductions in the fee paid to Uber? Tell me a little more about it.
            \item Tell me what it's like to do airport rides. How is it different compared to city rides? What incentivizes you to consider airport rides? How do you think the take rate varies for airport vs non airport rides?
        \end{itemize}
        \vspace*{\fill} %
    \end{minipage}
\end{figure}

\begin{figure}[h]
    
    \includegraphics[width=\linewidth]{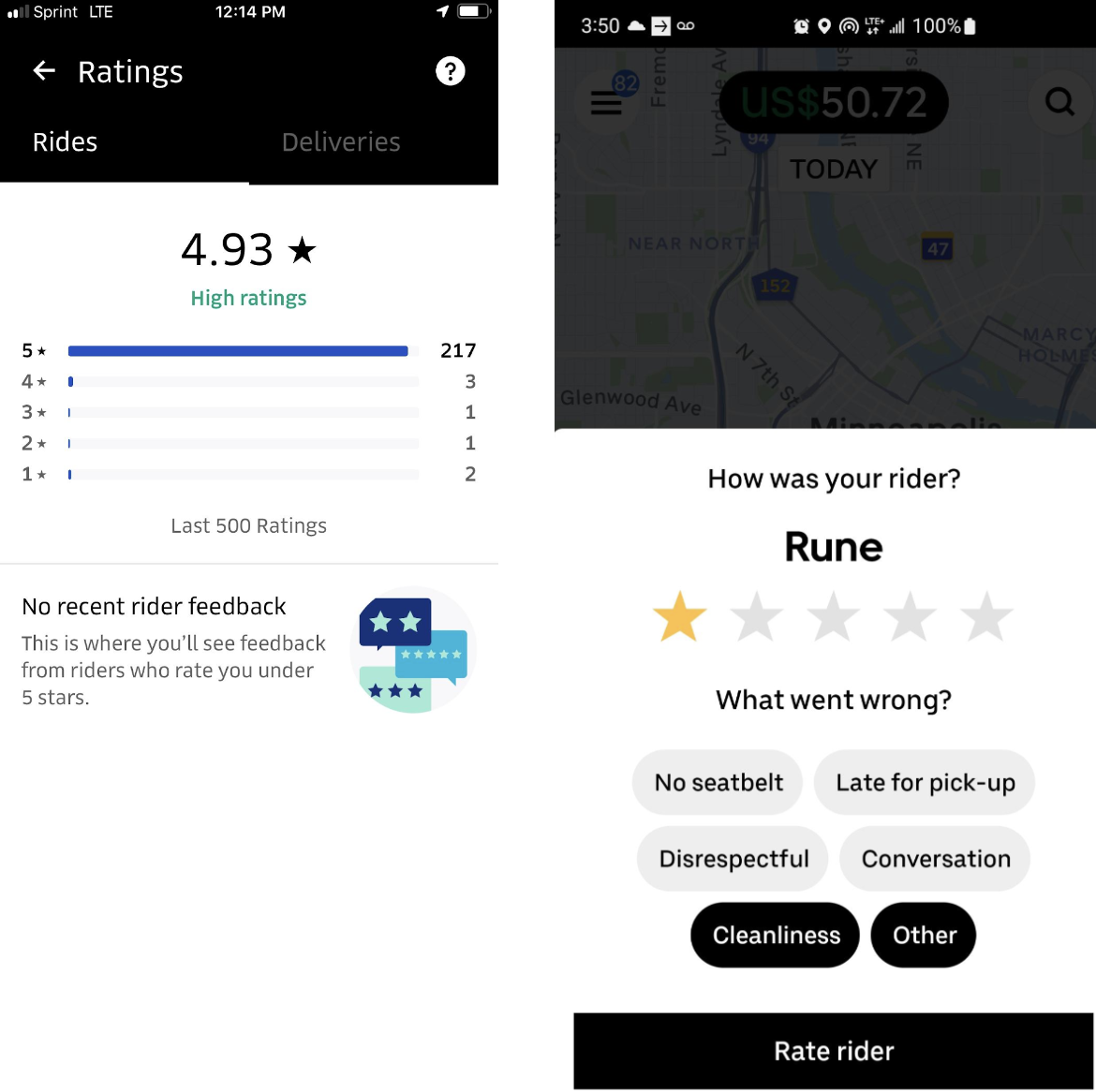} %
    \caption{Screenshot 8}
    \label{fig:screenshot8}  

    \bigskip %

    \begin{minipage}{\textwidth} %
    This screenshot is about driver and rider, ratings and reviews \\ \\
        \textbf{Probing Questions} \\
        \begin{itemize}
            \item How do you think about driver and rider ratings and reviews? 
            \item How does the driver and rider ratings and reviews influence your driving habits?
        \end{itemize}
    \end{minipage}
\end{figure}

\begin{figure}[h]
    \centering
    \begin{minipage}[c]{.5\textwidth}
        \centering
        \includegraphics[width=.9\linewidth]{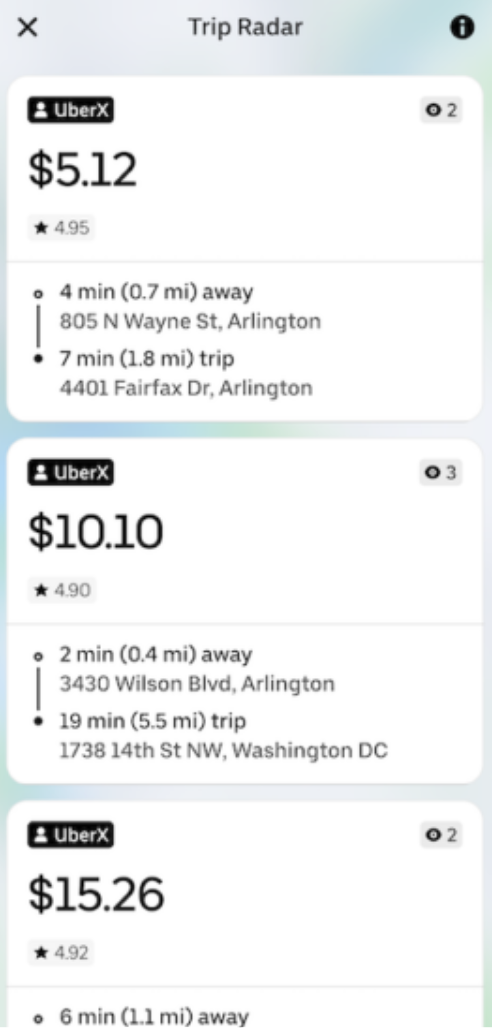} %
        \caption{Screenshot 9}
        \label{fig:screenshot9}
    \end{minipage}%
    \hspace{0.5cm}
    \begin{minipage}[c]{.4\textwidth}
        \vspace*{\fill} %
        This screenshot is about the trip radar. Trip radar is a feature that allows drivers to see and respond to all nearby product requests. When a driver declines a ride request, it gets relisted on the Trip Radar. The first driver to tap on the Trip Radar screen gets the ride. \\ \\
        \textbf{Probing Questions} \\ %
        \begin{itemize}
            \item Can you tell me a little more about how you understand and work with trip radar?
            \item What kind of ride requests do you think you get on trip radar?
            \item Trip Radar is a relatively new feature. How are you made aware of new features on the app?
        \end{itemize}
        \vspace*{\fill} %
    \end{minipage}
\end{figure}

\begin{figure}[h]
    \centering
    \begin{minipage}[c]{.5\textwidth}
        \centering
        \includegraphics[width=.9\linewidth]{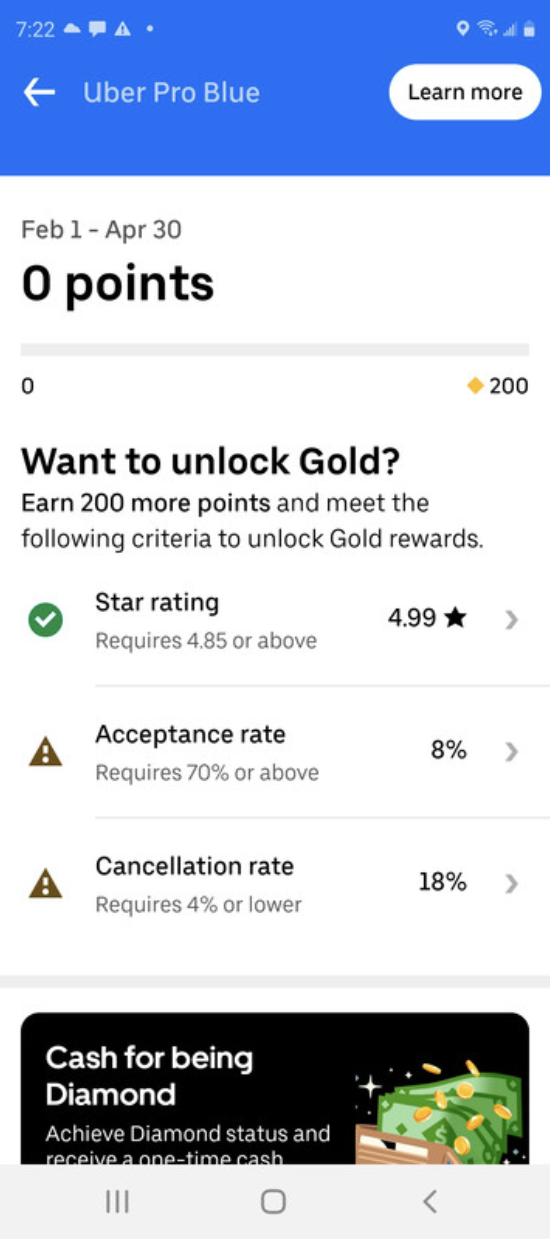} %
        \caption{Screenshot 10}
        \label{fig:screenshot10}
    \end{minipage}%
    \hspace{0.5cm}
    \begin{minipage}[c]{.4\textwidth}
        \vspace*{\fill} %
        This screenshot is about Uber Pro Rewards. \\ \\
        \textbf{Probing Questions} \\ %
        \begin{itemize}
            \item Can you tell me a little more about how you understand and work with Uber Pro rewards?
            \item How does Uber Pro Rewards influence your driving habits?
        \end{itemize}
        \vspace*{\fill} %
    \end{minipage}
\end{figure}

\clearpage

\section{LLM for Analysis of Public Forum Data, Prompt Engineering, Evaluation Methodology and Costs}

\subsection{Harnessing LLMs for Analysis of Online Public Forums' Data to Surface Harms and Concerns of a Community}
\label{asec:llm-background}
In recent years, data from online public forums, have been used for research to surface a multitude of harms and concerns facing various communities, drawing attention across multiple disciplines~\cite{proferes2021studying}. This data \footnote{available until at least December 2022 for Reddit; API access restrictions were enforced in February 2023; see \url{https://www.redditinc.com/policies/data-api-terms} and \url{https://pushshift.io/signup}} has been instrumental in studies ranging from healthcare~\cite{zamanifard2023social,xu2023technology, ho2017stigma, paxman2021everyone, hanna2016emoting, de2016stroke, faith2016exploring, prescott2017peer, wiggins2016stepping, parsons2019caregivers}, to political discourse and polarization \cite{papakyriakopoulos2023upvotes, de2021no, guimaraes2019analyzing, marjanovic2022quantifying, marchal2020polarizing}, to educational trends \cite{madsen2022communication, park2018harnessing}. In the context of gig economy research, studies have delved into rideshare workers' concerns about privacy, scams, and support systems \cite{rosenblat2016algorithmic, sannon2022privacy, yao2021together, watkins2022have, ma2018using}. However, these studies primarily relied on manual coding of small-scale datasets (the largest we found analyzed about 2.6K posts \cite{watkins2022have}). While detailed, this approach is limited in scalability due to the reliance on manual labeling of posts.

Recent advancements in Large Language Models (LLMs) offer an alternative for analyzing extensive online forum data addressing limitations of scalability. Inspired by Microsoft's \cite{nori2023can}, OpenAI's \cite{openai2023prompt} and other research \cite{deng2023llms} on prompt engineering, we developed a prompt to process over 1 million Reddit comments and submissions from r/uberdrivers and r/lyftdrivers (2019-2022). We leveraged LLMs to examine prevalent concerns about harms workers experience. Many gig workers turn to Reddit to voice complaints and concerns about the platform, making these forums a rich data source to learn about harms from workers' perspectives~\cite{watkins2022have}.

\subsection{Data Summary Statistics}
\label{asec:summary-stats}
We grouped submissions with their associated comments into batches of five to balance the LLM's error rate with computational cost, aiming to minimize data loss from potential failures while optimizing for execution time and monetary expenses. We encountered an 11\% error rate i.e. LLM failed to produce any output, due to various factors like throttling and content issues. The Generation phase thus resulted in 58,728 concerns related to the platform's AI and algorithmic decisions. These were then categorized into four themes in the Classification phase: Enhancing Transparency and Explainability (24,721 concerns, 42\%), Predictability and Worker Agency (12,728 concerns, 22\%), Better Safety and More Time (6,144 concerns, 10.5\%), and Ensuring Fairness and Non-Discrimination (4,280 concerns, 7\%). An additional 10,855 concerns (18.5\%) didn't fit into these themes and were categorized as ``Other''. Ultimately, after running through the Aggregation and Prevalence Prompts, and regrouping these concerns according to the four transparency-related codes generated from the interviews, we obtained a total of 31,340 concerns: promotions (11,490 concerns, 37\%), fares (3,546 concerns, 11\%), routes (9,083 concerns, 29\%), and task allocation (7,221 concerns, 23\%). Therefore, out of the 47,873 aggregated concerns, 16,533 concerns were either not relevant to the transparency codes or the secondary themes.

We provide summary statistics below for our Reddit dataset curated from  \texttt{r/uberdrivers} and \texttt{r/lyftdrivers} subreddits, for the time period 2019-2022.
\begin{table}[h]
\centering
\begin{tabular}{|l|c|}
\hline
\textbf{Statistic}                                  & \textbf{Value}    \\ \hline
Total number of posts                         & 65,377            \\ \hline
Total number of comments                            & 1,392,776         \\ \hline
Average number of comments per submission           & 21.30             \\ \hline
Average word count per post (excluding comments) & 82.04         \\ \hline
Average word count per post and its corresponding comments & 990.76 \\ \hline
Average number of words per comment                 & 32.47             \\ \hline
\end{tabular}
\caption{General Statistics of Posts and Comments}
\end{table}

\begin{table}[h]
\centering
\begin{tabular}{|l|c|c|}
\hline
\textbf{Statistic}                      & \textbf{r/UberDrivers} & \textbf{r/LyftDrivers} \\ \hline
Number of posts                   & 47,106                & 18,271                \\ \hline
Number of comments                      & 1,054,030             & 338,746               \\ \hline
\end{tabular}
\caption{Statistics for Specific Rideshare Platforms}
\end{table}

\begin{table}[htb]
\centering
\begin{tabular}{|l|c|}
\hline
\textbf{Statistic}                      & \textbf{Value}    \\ \hline
Number of groups (input to LLM)         & 13,077            \\ \hline
Number of concerns generated            & 58,728            \\ \hline
\end{tabular}
\caption{Group and Concern Statistics from 2019 to 2022}
\end{table}

\begin{table}[htb]
\centering
\begin{tabular}{|l|c|}
\hline
\textbf{Concern Category}                                  & \textbf{Number of Concerns} \\ \hline
Need for Enhancing Transparency and Explainability         & 24,721                      \\ \hline
Need for Greater Predictability and Worker Agency          & 12,728                      \\ \hline
Need for Ensuring Fairness and Non-Discrimination          & 4,280                       \\ \hline
Need for Better Safety and More Time                       & 6,144                       \\ \hline
Other Concerns                                             & 10,855                      \\ \hline
\end{tabular}
\caption{Categorized Concerns from 2019 to 2022}
\end{table}

\clearpage

\subsection{Prompts}
\label{asec:prompt}
\begin{lstlisting}[style=custombox, caption={GPT-4 Generation Prompt}]
Analyze a set of JSON objects, each representing a submission from the r/UberDrivers and r/LyftDrivers subreddits. For each JSON object, you will find the following information:

Submission Title: The title of the Reddit post.
Submission Body: The main content or message of the post.
Timestamp: The date and time when the post was submitted.
Group Key: A unique identifier that is common across all 5 submissions in the dataset.
Comments: A list of all comments made on the submission.

Generate a list of the most frequently occurring and impactful concerns due to AI and algorithmic platform features discussed by drivers within the input context.

Step 1: Identify mentions in submission bodies, titles, and/or comments about concerns that pertain to a lack of knowledge or available information regarding the platforms' algorithms for drivers, including but not limited to fares, routes, incentive programs, driver preferences, etc.
 
Step 2: Group similar concerns across comments and submissions to ensure a mutually exclusive list of concerns and avoid redundancy. For example, multiple mentions of fare calculation issues should be grouped under a single concern.

Step 3: From the grouped concerns, select the most representative quote for each concern. Ensure the quote clearly illustrates the specific concern due to AI and algorithmic platform features.

Step 4: Assess which concerns are mentioned most frequently and have the most significant impact on drivers.

Step 5: Create a list of these concerns in a JSON format. Each entry should include (with these specific field names):
"title": The title of the concern
"description": A brief description (10-20 words)
"quote": The selected representative quote.

Step 6: Ensure the final list is concise, precise, and specifically addresses drivers' concerns due to AI and algorithmic platform features present regarding the platform's algorithms and policies. Include only those concerns found in the input context without generalizing based on prior or outside knowledge. 

Step 7: Group any similar concerns to avoid redundancy. If there are no concerns, output ``No concerns''. Do not generate any other text.
\end{lstlisting}

\clearpage
\begin{lstlisting}[style=custombox, caption={GPT-4 Classification Prompt}]
Task: Analyze a list of rideshare drivers' transparency concerns. Each concern should be evaluated and categorized into one of the following five categories. For each concern listed, assign exactly one letter that corresponds to its most appropriate classification in the same order as the original list (preserving the serial number of the concern).

Categories:

A. Need for Enhancing Transparency and Explainability: Concerns due to the lack of information and explanations about AI and algorithmic features, required for drivers to do their work. These include opaque trip details, unclear surge boundaries, bonus and quest clarity issues, lack of wage breakdown, platform take rate and clarity on how prices and bonuses are calculated, and the impact of ratings on metrics.

B. Need for Greater Predictability and Worker Agency: Concerns due to significant variation in AI and algorithmic features leading to reduced worker agency and diminished predictability of work conditions. These include unpredictable wages, misleading destination filter, immense variation in surge prices and quest matches across drivers and location, and deceptive trip offers.

C. Need for Better Safety and More Time: Concerns focused on safety risks and time pressures due to the AI and algorithmic features. These include dangerous multitasking, compromised route safety, unattainable quests, and acceptance rate concerns, emphasizing the need for improved safety and time management in app design.

D. Need for Ensuring Fairness and Non-Discrimination: Concerns related to algorithmic wage discrimination. These include unequal pay for similar work, earnings below the prevailing minimum wage, influence of the demographic characteristics of drivers and riders on earnings, all caused by opaque AI and algorithmic features.

E. Other Concerns: Any concerns that do not fit into the above categories.

Output Format: Present the analysis in a dictionary format with the serial number of the concern as the key and the classification as the value, preserving the order of the concerns.

{1: A, 2: B, 3: C}

Note: Ensure that each concern is classified under only one of the aforementioned categories and that there is one classification corresponding to each concern in the input, e.g., for 400 transparency concerns, there should be 400 items in the dictionary. Do not generate any other text. 
\end{lstlisting}

\clearpage
\begin{lstlisting}[style=custombox, caption={GPT-4 Aggregation and Ranking Prompt}]
The data contains a list of concerns of rideshare drivers obtained from discussions on the r/uberdrivers and r/lyftdrivers subreddits. Each concern title and its description is present on consecutive lines.

These concerns are relevant to <insert category description from Classification Prompt>

Identify the 5 most frequently occurring themes of concerns. Ensure each concern is sufficiently different from the others on the list. Don't repeat the same concern in the list. If there are similar concerns, group them and find another one. 

Provide the output in a rank ordered format:
{concern_rank: <a number between 1-5>, concern_title: string, concern_description: string of 10-20 words}

\end{lstlisting}

\begin{lstlisting}[style=custombox, caption={GPT-4 Prevalence Prompt}]
Classify each line which contains the title and description of a concern from rideshare drivers on Reddit into the following 6 categories:

<list the 5 categories A to E from the aggregation prompt output>
F: Other

Output Format: Present the analysis in a dictionary format with the serial number of the concern as the key and the classification as the value, preserving the order of the concerns.

{1: A, 2: B, 3: C}

Note: Ensure that each concern is classified under only one of the aforementioned categories and that there is one classification corresponding to each concern in the input, e.g., for 400 concerns, there should be 400 items in the dictionary. Do not generate any other text.
\end{lstlisting}

\subsection{Evaluation}
\label{asec:evaluation}
\subsubsection{Methodology}
To evaluate the LLM outputs we propose a mix of human and computational evaluation methods based on the following metrics: \textit{factuality} and \textit{completeness} for generation, \textit{accuracy} for classification and prevalence, and \textit{distinctness} and \textit{coverage} for aggregation prompts, respectively.

\noindent \textbf{Generation Prompt:}\\
    \begin{itemize}
        \item \textit{Factuality:}  Determines if the LLM's output concerns reflect the source data. Evaluators respond to: \textit{``Is this candidate concern (and associated quote) output by the LLM present in the reference human generated list of transparency concerns? Answer Yes or No''}. The factuality score is the proportion of accurate LLM-generated concerns, akin to precision.
        \item \textit{Completeness:}  Evaluates the LLM's ability to capture all relevant concerns, asking: \textit{``Is this reference human concern present in the candidate concern (and associated quote) output by the LLM? Answer Yes or No.''} The completeness score denotes the proportion of human-identified concerns also recognized by the LLM, akin to recall.
    \end{itemize}
\noindent \textbf{Classification and Prevalence Prompts:} \begin{itemize}
    \item \textit{Accuracy:} Assesses the correctness of LLM classifications, measuring the proportion of accurately classified samples.
\end{itemize}
\noindent \textbf{Aggregation Prompt:} \begin{itemize}
    \item \textit{Distinctness:} We measure the \textit{proportion of unique most-similar topics associated with each secondary-theme across all the secondary-themes}. A higher value indicates a greater diversity in the subthemes.
    \item \textit{Coverage(k):} We measure the proportion of unique most-similar topics associated with each of the `n' secondary-themes that are also found among the most frequent `nk' topics of the overall text. For example, if n=5 and k=2, we would look at the 5 secondary-themes amongst the top 10 most frequent overall topics. A higher Coverage(k) indicates that the secondary-themes represent prevalent or significant themes in the text.
\end{itemize} 

\subsubsection{Results:} We present the results of our evaluation below.\\

\noindent \textbf{Generation Prompt.}  We randomly sampled 125 submissions and all the 2,511 associated comments, jointly annotated by two researchers, to create a reference list of concerns. The two researchers then jointly evaluated the LLM outputs against the reference list of concerns, and resolved any differences through discussions. We obtained a factuality score of 0.55 and a completeness score of 0.78. Scores are statistically significantly different from chance measured via a binomial test (p-value < 0.05). Lower factuality was primarily due to the LLM identifying concerns part of the input context, but not directly related to AI and algorithmic decisions. Moreover, the lower factuality score was less concerning as outputs not factual (i.e., identified by the LLM but not by humans) were mostly filtered out into the ``Other'' category during the classification step and excluded from any further analysis. \\
    
\noindent \textbf{Classification and Prevalence Prompts.}  Two researchers annotated 100 LLM-classified concerns to establish ground truth labels, assigning one label per concern. They compared LLM outputs with these labels, achieving a consensus accuracy of 0.74. The achieved accuracy is significant because misclassified samples, excluding those labeled as ``Other,'' can still contribute to the analysis. This is because concerns might fit into multiple categories depending on the analytical perspective. Even if placed in a different category, these concerns remain valuable for later aggregation and prevalence prompts. Using a similar methodology, researchers obtained an accuracy of 0.82 for the prevalence prompts' output. Scores are statistically significantly different from chance measured via a binomial test (p-value < 0.05). \\
    
\noindent \textbf{Aggregation Prompt.} We employed BERTopic for topic modeling (see Appendix \ref{asec:bertopic-hyperparams} for hyperparameter details). Given the large volume of concerns (4K-25K) per primary theme, human evaluation on a subset would not accurately capture the task's nuances. Therefore, we adopted this well-established computational method for evaluation. We fit a topic model on the entire text (title+description) associated with each primary theme to obtain the top 5 and 10 topics. Subsequently, we identified the most similar topic for each of the 5 sub-themes. The results demonstrated a distinctness of 0.80, coverage(1) of 0.95, and coverage(2) of 1.00, indicating that the LLMs' outputs were both distinct and well-represented among the most frequent topics.  Scores are statistically significantly different from chance measured via a binomial test (p-value < 0.05).

\subsection{Financial Expenditure of Using Large Language Models (LLMs)}
\label{asec:costs}

\begin{table}[ht]
\centering
\begin{tabular}{ccc}
\toprule
\textbf{Cost Category} & {\textbf{Token Quantity}} & {\textbf{Cost (USD)}} \\
\midrule
Input Token Rate & {-} & {\$0.01 per 1K tokens} \\
Output Token Rate & {-} & {\$0.03 per 1K tokens} \\
\addlinespace
Total Input Tokens & 135.12 million & \$1351.20 \\
Total Output Tokens & 10.37 million & \$311.10 \\
\addlinespace
\midrule
\textbf{Total Expenditure} & {-} & \textbf{\$1662.30} \\
\bottomrule
\end{tabular}
\caption{GPT4-Turbo Token Usage and Associated Costs}
\label{tab:llm_costs}
\end{table}

In this section, we delve into the costs incurred from utilizing the GPT4-Turbo LLM to synthesize concerns from the Reddit data. This model operates on a token-based system, where both the input (data fed into the model) and output (data generated by the model) are quantified in ``tokens''. A token in this context can be understood as a piece of text, which could be as small as a single character or as large as a word. The exact size of a token varies based on the complexity of the language and the specific processing needs of the model. To make this more relatable, consider that an average English word is approximately equivalent to four tokens. This conversion isn't exact due to variations in word length and language complexity, but it provides a general idea of how token counts translate into the more familiar concept of word counts. The pricing model for using GPT4-Turbo is based on the number of these tokens processed. Therefore, the more text (or tokens) we input into or receive from the model, the higher the cost. 

The access to the LLM and associated compute resources through Microsoft Azure was jointly funded through a partnership between our university and Microsoft. 
The total expenditure for the GPT4-Turbo usage was \$1,662.30. This funding enabled us to leverage the capabilities of GPT4-Turbo to successfully meet our research objectives.

\clearpage
\subsection{BERTopic Hyperparameters}
\label{asec:bertopic-hyperparams}
\begin{lstlisting}
# Load CSV data
df = pd.read_csv(filename)

# Concatenate title and description into one text column
df['text'] = df['title'] + " " + df['description']

dim_model = UMAP(n_neighbors=10,
                 n_components=10,
                 min_dist=0.0,
                 metric='cosine',
                 random_state=100)

# Initialize BERTopic
topic_model = BERTopic(
    nr_topics=10,
    min_topic_size=85,
    n_gram_range=(1, 2),
    umap_model=dim_model
)

# Fit BERTopic on the concatenated texts from the CSV
topic_model.fit(df['text'])
\end{lstlisting}

\section{Transparency Report Indicators}
\begin{figure}[!htb]
    \centering
    \includegraphics[width=\columnwidth]{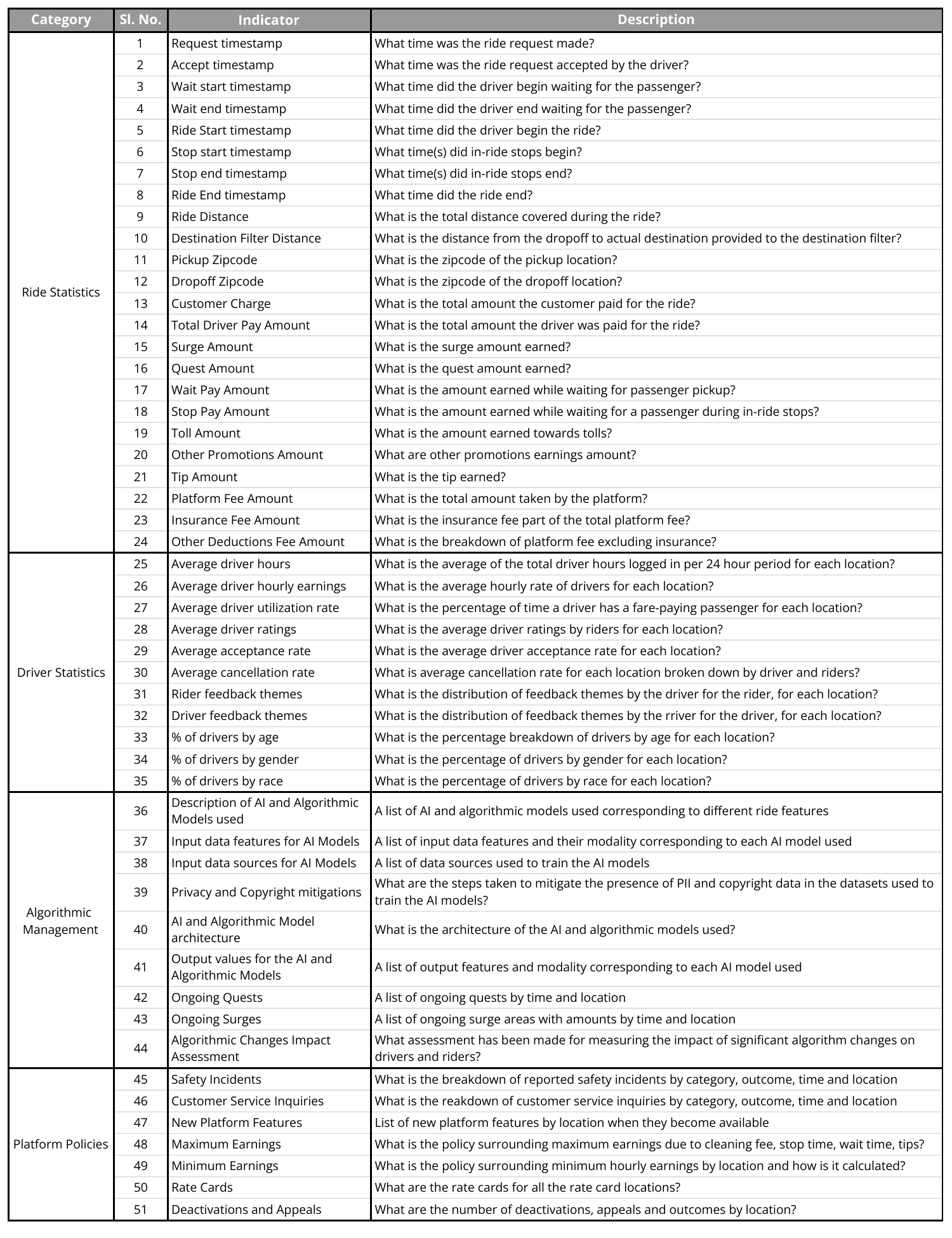}
    \caption{Indicators for the Rideshare Transparency Report}
    \label{fig:transparency-report}
\end{figure}

We provide a list of transparency indicators in Figure \ref{fig:transparency-report}. 
\clearpage

\end{document}